\documentclass[aps,prd,preprintnumbers,showpacs,superscriptaddress,nofootinbib,amsmath,amssymb,floats,floatfix,showkeys,notitlepage,longbibliography]{revtex4-1}
\usepackage{comment}
\usepackage{graphicx}
\usepackage{subfigure}
\usepackage{palatino}
\usepackage[commandnameprefix=always]{changes}
\usepackage{hyperref}
\hypersetup{colorlinks=true,linkcolor=red,urlcolor=red,citecolor=red}
\usepackage[toc,page]{appendix}
\usepackage[normalem]{ulem}

\usepackage{lipsum}
\usepackage{graphicx}
\usepackage{subfigure}
\usepackage{palatino}
\usepackage{float}
\usepackage{sans}
\usepackage{adjustbox}
\usepackage{latexsym}
\usepackage{amsmath}
\usepackage{amssymb}
\usepackage{amsfonts}
\usepackage{dcolumn}
\usepackage{bm}
\usepackage{tikz}
\usepackage{bigints}
\usepackage{array,tabularx,multirow,booktabs}
\usepackage[tracking=true]{microtype}
\SetTracking{}{500}
\SetTracking{encoding={*}, shape=sc}{40}
\allowdisplaybreaks
\usepackage{adjustbox}
\usepackage{latexsym}
\usepackage{amsmath}
\usepackage{amssymb}
\usepackage{amsfonts}
\usepackage{dcolumn}
\usepackage{bm}
\usepackage{tikz}
\usepackage{bigints}
\usepackage{array,tabularx,multirow,booktabs}
\usepackage[tracking=true]{microtype}
\usepackage{color}
\allowdisplaybreaks

\begin{document}

\title{Thermodynamic Topology and Phase Space Analysis of AdS Black Holes Through Non-Extensive Entropy Perspectives}

\author{Saeed Noori Gashti
}
\email{saeed.noorigashti@stu.umz.ac.ir}
\affiliation{School of Physics, Damghan University,\\ P. O. Box 3671641167, Damghan, Iran}

\author{Behnam Pourhassan}
\email{b.pourhassan@du.ac.ir}
\affiliation{School of Physics, Damghan University, Damghan 3671645667, Iran.}
\affiliation{Center for Theoretical Physics, Khazar University, 41 Mehseti Street, Baku, AZ1096, Azerbaijan}
\affiliation{Physics Department, Istanbul Technical University,
Istanbul 34469, Turkey}
\affiliation{University Centre for Research \& Development, Chandigarh University, Mohali, Punjab, 140413, India}

\author{\.{I}zzet Sakall{\i}
}
\email{izzet.sakallı@emu.edu.tr}
\affiliation{Physics Department, Eastern Mediterranean
University, Famagusta, 99628 North Cyprus, via Mersin 10, Turkiye.}

\begin{abstract}
In this paper, we study the thermodynamic topology of AdS Einstein-power-Yang-Mills black holes, examining them through both the bulk-boundary and restricted phase space (RPS) frameworks. We consider various non-extensive entropy models, including Barrow ($\delta$), Rényi ($\lambda$), Sharma-Mittal ($\beta$, $\alpha$), Kaniadakis ($K$), and Tsallis-Cirto entropy ($\Delta$). Initially, we analyze the thermodynamic topology within the bulk-boundary framework. Our findings highlight the influence of free parameters on topological charges. We observe two topological charges $(\omega = +1, -1)$  with respect to the non-extensive Barrow parameter and also with ($\delta=0$) in Bekenstein-Hawking entropy. For Rényi entropy, different topological charges are observed depending on the value of the $\lambda$ with a notable transition from three topological charges $(\omega = +1, -1, +1)$ to a single topological charge $(\omega = +1)$ as $\lambda$ increases. Also, by setting $\lambda$ to zero results in two topological charges $(\omega = +1, -1)$. Sharma-Mittal entropy exhibits three distinct ranges of topological charges influenced by the $\alpha$ and $\beta$ with different classifications viz $\beta$ exceeds $\alpha$, we will have $(\omega = +1, -1, +1)$, $\beta = \alpha$,  we have $(\omega = +1, -1)$ and for $\alpha$ exceeds $\beta$ we face $(\omega = -1)$. Also, Kaniadakis entropy shows variations in topological charges viz we observe $(\omega = +1, -1)$ for any acceptable value of $K$, except when $K = 0$, where a single topological charge $(\omega = -1)$. In the case of Tsallis-Cirto entropy, for small parameter $\Delta$ values, we have $(\omega = +1)$ and when $\Delta$ increases to 0.9, we will have $(\omega = +1, -1)$. A particularly intriguing aspect of this research is its application to the RPS framework. When we extend our analysis to this space using the specified entropies, we find that the topological charge consistently remains $(\omega = +1)$ independent of the specific values of the free parameters for Rényi, Sharma-Mittal, and Tsallis-Cirto. Additionally, for Barrow entropy in RPS, when $\delta$ increases from 0 to 0.8, the number of topological charges rises. Finally for Kaniadakis entropy, at small values of $K$, we observe $(\omega = +1)$. However, as the non-extensive parameter $K$ increases, we encounter different topological charges and classifications with  $(\omega = +1, -1)$.
\end{abstract}

\date{\today}



\maketitle
\tableofcontents
\section{Introduction}
Stephen Hawking's area theorem \cite{900} posits that the total horizon area of black holes cannot decrease over time during any physical process that adheres to classical physics laws. This theorem suggests that black holes possess thermodynamic properties, behaving similarly to entropy in thermodynamic systems. Jacob Bekenstein expanded on this concept by proposing that a black hole's entropy is proportional to the area of its event horizon. This relationship, known as Bekenstein-Hawking entropy, highlights a profound connection between black hole geometry and thermodynamic entropy \cite{901,902}. The analogy between black hole thermodynamics and classical thermodynamics was further reinforced by Stephen Hawking's discovery of Hawking radiation. This phenomenon, driven by quantum effects near the event horizon, causes black holes to emit thermal radiation. Consequently, black holes can be assigned a temperature, known as Hawking temperature, which is inversely proportional to their mass \cite{903,904,905}.

A novel method examines the topological charge of black holes by interpreting black hole solutions as topological defects within the thermodynamic parameter space. Using the generalized off-shell free energy, black holes are classified based on their topological charge, determined by the winding numbers of these defects. Black holes with positive winding numbers are considered locally stable, while those with negative winding numbers are deemed locally unstable. This innovative approach provides new insights into the thermodynamic stability of black holes and offers valuable perspectives on phase transitions and critical phenomena in black hole thermodynamics. It has been applied to various black holes, including those in anti-de Sitter (AdS) spacetime, revealing new types of critical points and phase behaviors \cite{a19,a20}.

One effective approach to gaining a topological perspective in thermodynamics is to utilize Duan’s topological current $phi$- mapping theory. Wei et al. introduced two distinct methods to explore topological thermodynamics, focusing on temperature and generalized free energy functions \cite{a19,a20}. The first method involves analyzing the temperature function by eliminating pressure and employing the auxiliary and topological parameters $ ( \frac{1}{\sin \theta} )$. Based on these assumptions, a potential is constructed. The second method assumes that black holes can be considered defects within the thermodynamic parameter space. Their solutions are examined using the generalized off-shell free energy. In this context, the stability and instability of the resulting BH solutions are indicated by positive and negative winding numbers, respectively. Additionally, in this theory, the characteristics of a field configuration are determined by the zero points of the field in space.

The topological method for black hole thermodynamics has gained popularity due to its simplicity in examining thermodynamic properties. It has been used to investigate the Hawking-Page phase transition of Schwarzschild-AdS black holes and their holographic counterparts, which relate to the confinement-deconfinement transition in gauge theories. Quantum gravity corrections, expressed through higher-derivative terms, have been studied for black holes in Einstein-Gauss-Bonnet and Lovelock gravity. These corrections provide insights into the behavior of black holes in higher-dimensional spacetimes and the effects of quantum gravity. Although these studies mainly focus on static black holes, the topological approach has also been extended to rotating black holes, offering significant insights into their thermodynamic properties, stability, topological classification and topological photon spheres \cite{20a,21a,22a,23,24,25,26,27,28,29,31,33,34,35,37,38,38a,38b,38c,39,40,41,42,43,44,44c,44d, 44d'}.

In this pper, we aim to explore the topology of holographic thermodynamics of AdS Einstein-power-Yang-Mills black holes using non-extensive entropies such as Barrow, Rényi, Sharma-Mittal, Kaniadakis and Tsalis Cirto entropy. Our objective is to identify the topological class of these black holes and compare it with the Bekenstein-Hawking entropy. Non-extensive entropy, often associated with Tsallis entropy, is a generalization of the traditional Boltzmann-Gibbs entropy. This concept was introduced by Tsallis to address systems where the conventional assumptions of extensive entropy do not apply. In classical thermodynamics, entropy is extensive, meaning it scales linearly with the system's size. However, many physical systems exhibit non-extensive behavior due to long-range interactions, fractal structures, or other complexities \cite{aa,bb,cc,dd,ee,ff}. Non-extensive entropy has been applied to various astrophysical phenomena, including the distribution of stellar objects and the dynamics of galaxy clusters. It helps model systems where gravitational interactions are long-range and cannot be described by extensive entropy. Non-extensive entropy extends information theory concepts to systems with non-standard probability distributions. It is used in coding theory, data compression, and the analysis of complex networks \cite{aa,bb,cc,dd,ee,ff}.

Holographic thermodynamics applies the principles of holography to the study of black hole thermodynamics. This approach often involves the AdS/CFT correspondence, which posits a relationship between a gravitational theory in an anti-de Sitter (AdS) space and a conformal field theory (CFT) on its boundary. This duality allows physicists to study complex gravitational systems using the simpler, well-understood properties of quantum field theories. Thus, one can study two spaces with features such as bulk-boundary correspondence and restricted phase space. Bulk-boundary correspondence connects the properties of a bulk system (like a black hole in AdS space) with those of its boundary (the CFT). This principle is crucial in understanding topological phases of matter and has applications in condensed matter physics and high-energy physics. It essentially states that the behavior of a system's boundary can reveal information about the bulk properties. Restricted phase space thermodynamics is a newer formalism that modifies traditional black hole thermodynamics by fixing certain parameters, such as the AdS radius, as constants. This approach eliminates the need for pressure and volume as thermodynamic variables, instead using the central charge and chemical potential. This formalism maintains the Euler relation equation, providing a consistent framework for studying black hole thermodynamics \cite{701,702,703,704,705,706,707,708,709,710,711,712,713,714,715,716,717}.

So, with respect to the above concepts, we organize the paper as follows:\\
In section 2 we will delve into the concept of Nonextensive Entropy, covering models, formulas, and their applications in various physical systems. This section will overview Nonextensive Entropy, associated with Barrow, Rényi, Sharma-Mittal, Kaniadakis, and Tssalis Cirto, extending the traditional Boltzmann-Gibbs framework to accommodate systems with long-range interactions, fractal structures, and other complexities that exhibit non-extensive behavior. In section 3 we will explain the thermodynamic topology using the generalized Helmholtz free energy method. This section will discuss how this method allows us to classify black holes based on their topological charge, determined by the winding numbers of topological defects in the thermodynamic parameter space. We will also overview the implications of this classification for understanding the stability and phase transitions of black holes.
In section 4 we will provide a comprehensive overview of the black hole model within the frameworks of bulk-boundary correspondence and restricted phase space. This section will include detailed calculations and discussions on the thermodynamic topology of the model, focusing on non-extensive entropies such as Barrow, Rényi, Sharma-Mittal, Kaniadakis, and Tssalis Cirto entropy. We will examine how these entropies influence the thermodynamic properties and stability of black holes and compare them with the traditional Bekenstein-Hawking entropy. In section 5 we will present our conclusions and summarize the key findings of our study. This section will reflect on the insights gained from our exploration of non-extensive entropies and thermodynamic topology, discussing the broader implications of our results for the field of black hole thermodynamics. We will also suggest potential directions for future research, building on the foundations laid by our work.
\section{Non-extensive Entropy}
We begin with the standard thermodynamical entropy in black hole physics, a significant concept that led to the development of black hole thermodynamics. The Bekenstein-Hawking entropy is expressed as \cite{N1},
\begin{equation}\label{N1}
\begin{split}
S = \frac{A}{4G},
\end{split}
\end{equation}
where $(A \equiv 4\pi r_h^2)$ is the area of the horizon, and $(r_h)$ is the horizon radius (using the areal radius as the radial coordinate). This formula links the entropy of a black hole to its surface area, providing a profound connection between thermodynamics and general relativity.
\subsection{Barrow Entropy}
However, this proposal is not unique. Depending on the system under consideration, different entropies may be introduced. Here the Barrow entropy concepts proposed as \cite{N2,N3},
\begin{equation}\label{N2}
\begin{split}
S_B = \left(\frac{A}{A_{Pl}}\right)^{1 + \delta/2},
\end{split}
\end{equation}
where $(A)$ is the usual black hole horizon area, and $(A_{Pl} \equiv 4G)$ is the Planck area. The Barrow entropy resembles the Tsallis non-extensive entropy but is based on different physical principles. It was proposed as a model for the effects of quantum gravitational spacetime foam, with the deformation quantified by the exponent $(\Delta)$. The Barrow entropy reduces to the standard Bekenstein-Hawking entropy in the limit $(\Delta \to 0)$, while $(\Delta = 1)$ corresponds to maximal deformation. This concept is particularly intriguing as it incorporates potential quantum gravitational effects, offering insights into the nature of spacetime at the smallest scales.
\subsection{Rényi Entropy}
Rényi entropy is defined as \cite{N4,N5,N6},
\begin{equation}\label{N3}
\begin{split}
S_R = \frac{1}{\lambda} \ln(1 + \lambda S),
\end{split}
\end{equation}
where $S$ is identified with the Bekenstein-Hawking entropy, and $(\lambda)$ is a parameter. The Rényi entropy was initially proposed as an index specifying the amount of information, without any relation to the statistics of physical systems. It provides a generalized measure of entropy that can interpolate between different entropy definitions, making it versatile for various applications in information theory and statistical mechanics.
\subsection{Sharma-Mittal Entropy}
Sharma-Mittal entropy is given by \cite{N7,N8,N9},
\begin{equation}\label{N4}
\begin{split}
S_{SM} = \frac{1}{\alpha} \left[(1 + \delta S_T)^{\alpha/\beta} - 1\right],
\end{split}
\end{equation}
where $(S_T)$ is the Tsallis entropy, and $\alpha$ and $(\beta)$ are free phenomenological parameters determined by the best fit to experimental data. The Sharma-Mittal entropy can be seen as a combination of the Rényi and Tsallis entropies. It offers a flexible framework for modeling systems with varying degrees of complexity and interaction strengths.
\subsection{Kaniadakis Entropy}
Kaniadakis entropy is given by \cite{N10},
\begin{equation}\label{N5}
\begin{split}
S_K = \frac{1}{K} \sinh(KS),
\end{split}
\end{equation}
which reproduces the Bekenstein-Hawking entropy in the limit $(K \to 0)$. It can be regarded as a generalization of the Boltzmann-Gibbs entropy arising in relativistic statistical systems. Kaniadakis entropy is useful in describing systems where relativistic effects are significant, providing a more accurate description of entropy in high-energy physics and cosmology.
These various entropy concepts highlight the diverse approaches to understanding entropy in different physical contexts. Each entropy measure offers unique insights and tools for tackling the complexities of different physical systems, from black holes to quantum gravity and beyond.
\subsection{Tsallis-Cirto Entropy}
The Tsallis-Cirto entropy is a generalization of the traditional Bekenstein-Hawking entropy, incorporating principles from non-extensive statistical mechanics. This entropy measure is particularly useful in cosmology for exploring the thermodynamic properties of gravitational systems, such as black holes and the universe itself. In cosmology, the Tsallis-Cirto entropy has been applied to derive modified Friedmann equations that describe the universe’s expansion. These modifications can provide insights into dark energy and the universe’s accelerated expansion. One interesting application is in the context of entropic cosmology, where gravity is considered an emergent phenomenon resulting from the statistical behavior of microscopic degrees of freedom. This approach can lead to new perspectives on the nature of dark energy and the dynamics of the universe. For the Tsallis-Cirto proposal, the entropy is modified as \cite{N11},
\begin{equation}\label{N6}
\begin{split}
S_{TC} = (S_{BH})^\Delta,
\end{split}
\end{equation}
where $(S_{BH})$ is the Bekenstein-Hawking entropy. When $(\Delta \to 1)$, this reduces to the standard entropy of a black hole. This modification allows for a broader understanding of entropy in various cosmological contexts, potentially offering new insights into the fundamental nature of the universe.
\section{Thermodynamic Topology}
Recent advancements have introduced innovative methods for analyzing and computing critical points and phase transitions in black hole thermodynamics. One prominent approach is the topological method, which leverages Duan’s topological current $\phi$-mapping theory to adopt a topological perspective in thermodynamics \cite{a19,a20}. This topological approach provides a robust framework for understanding the stability and phase transitions of black holes, offering new insights into their thermodynamic behavior. The generalized free energy is given by \cite{a19,a20},
\begin{equation}\label{F1}
\begin{split}
\mathcal{F} = M - \frac{S}{\tau},
\end{split}
\end{equation}
where $\tau$ signifies the Euclidean time period. A vector $\phi$ is constructed with components derived from the partial derivatives as follows,
\begin{equation}\label{F2}
\begin{split}
\phi = \left(\frac{\partial \mathcal{F}}{\partial r_H}, -\cot \Theta \csc \Theta \right).
\end{split}
\end{equation}
In this scenario, $\phi^\Theta$ becomes infinite, and the vector points outward at the angles $\Theta = 0$ and $\Theta = \pi$. By applying Duan's $\phi$-mapping topological current theory, we can define a topological current as follows,
\begin{equation}\label{F3}
\begin{split}
j^\mu = \frac{1}{2\pi} \varepsilon^{\mu\nu\rho} \varepsilon_{ab} \partial_\nu n^a \partial_\rho n^b, \quad \mu, \nu, \rho = 0, 1, 2,
\end{split}
\end{equation}
where $n$ is defined as $(n^1, n^2)$, with $n^1 = \frac{\phi^r}{|\phi|}$ and $n^2 = \frac{\phi^\Theta}{|\phi|}$. The topological number or total charge $W$ can be determined as follows,
\begin{equation}\label{F4}
\begin{split}
W = \int_\Sigma j^0 d^2 x = \sum_{i=1}^n \beta_i \eta_i = \sum_{i=1}^n \omega_i,
\end{split}
\end{equation}
where, $\beta_i$ represents the positive Hopf index, and $\eta_i$ is defined as the sign of $j^0(\phi/x)_{z_i}$, which can be either +1 or -1. The term $\omega_i$ denotes the winding number associated with the $i$-th zero point of $\phi$ within the region $\Sigma$.
\section{AdS Einstein-power-Yang-Mills black holes}

AdS Einstein-power-Yang-Mills (EPYM) black holes are unique solutions in higher-dimensional gravity theories that feature a non-linear extension of the Yang-Mills field. This field, a gauge field, describes the interactions of elementary particles within quantum field theory. The power-Yang-Mills field is characterized by a Lagrangian density that incorporates a power of the Yang-Mills field strength tensor. These black holes are solutions to the Einstein field equations with a negative cosmological constant and a power-Yang-Mills source term. AdS EPYM black holes possess intriguing characteristics such as horizon structure, thermodynamic behavior, phase transitions, and Joule-Thomson expansion. The metric for an AdS EPYM black hole is described by \cite{M1,M2},
\begin{equation}\label{M1}
\begin{split}
ds^2 = -f(r)dt^2 + \frac{dr^2}{f(r)} + r^2 d\Omega_2^2,
\end{split}
\end{equation}
where,
\begin{equation}\label{M2}
\begin{split}
f(r) = 1 - \frac{2M}{r} + \frac{r^2}{l^2} + \frac{(2q^2)^\gamma}{2(4\gamma - 3)r^{4\gamma - 2}},
\end{split}
\end{equation}
In the Eq.(\ref{M2}), $M$ denotes the mass, $q$ is the Yang-Mills charge, and $\gamma$ determine the non-linear YM charge parameter.  The temperature of an AdS EPYM black hole is given by,
\begin{equation}\label{M3}
\begin{split}
T = \frac{1}{4\pi r_h} \left(1 + 8\pi P r_h^2 - \frac{(2q^2)^\gamma}{2r_h^{4\gamma - 2}}\right),
\end{split}
\end{equation}
where $r_h$ is the radius of the event horizon. The entropy of an AdS EPYM black hole is as $S = \frac{A}{4}.$ where $A$ is the area of the horizon. This entropy quantifies the disorder or information loss associated with the black hole, based solely on the horizon area according to the Bekenstein-Hawking formula.
\subsection{Bulk boundary thermodynamics}
Based on Eq.(\ref{M2}), the function $f(r)$ for AdS EPYM black holes is reformulated as,
\begin{equation}\label{BB1}
\begin{split}
f(r) = 1 - \frac{2GM}{r} + \frac{r^2}{l^2} + \frac{G(2q^2)^\gamma}{2(4\gamma - 3)r^{4\gamma - 2}}
\end{split}
\end{equation}
The thermodynamic properties, including the Hawking temperature, mass, and Helmholtz free energy for AdS EPYM black holes, are described by the following equations. So, the Hawking temperature is given by,
\begin{equation}\label{BB2}
\begin{split}
T = \frac{1 + 8\pi G P r_h^2 - \frac{(2q^2)^\gamma G}{2r_h^{4\gamma - 2}}}{4\pi r_h}.
\end{split}
\end{equation}
The mass $M$ is expressed as,
\begin{equation}\label{BB3}
\begin{split}
M = \frac{r_h \left(1 + 8\pi GP r_h^2 - \frac{q^{2\gamma} 2^{1-\gamma} G}{r_h^{4\gamma - 2} (4\gamma - 3)}\right)}{2G}.
\end{split}
\end{equation}
The constant $G$ is calculated as,
\begin{equation}\label{BB4}
\begin{split}
G = 2 \left(8 (q^2)^\gamma r_h^{-4\gamma + 2} \gamma2^{- 1 + \gamma} + 16\pi P r_h^2 - (q^2)^\gamma r_h^{-4\gamma + 2} 2\gamma\right)^{-1}
\end{split}
\end{equation}
\subsubsection{Barrow entropy, bulk boundary thermodynamics and thermodynamic topology}
\begin{figure}[h!]
 \begin{center}
 \subfigure[]{
 \includegraphics[height=4cm,width=4cm]{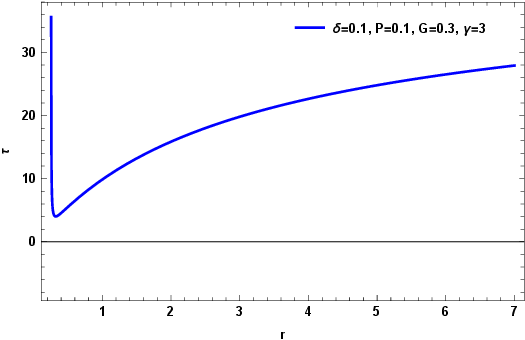}
 \label{1a}}
 \subfigure[]{
 \includegraphics[height=4cm,width=4cm]{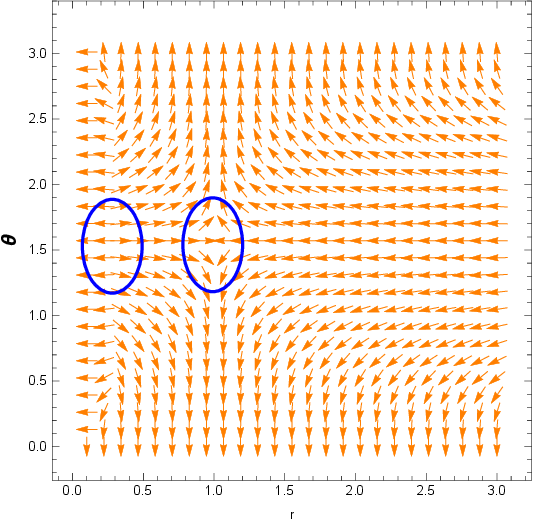}
 \label{1b}}
 \subfigure[]{
 \includegraphics[height=4cm,width=4cm]{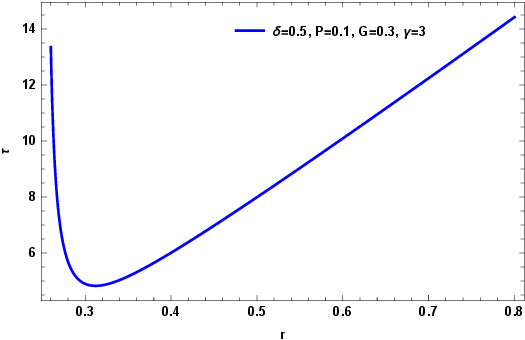}
 \label{1c}}
 \subfigure[]{
 \includegraphics[height=4cm,width=4cm]{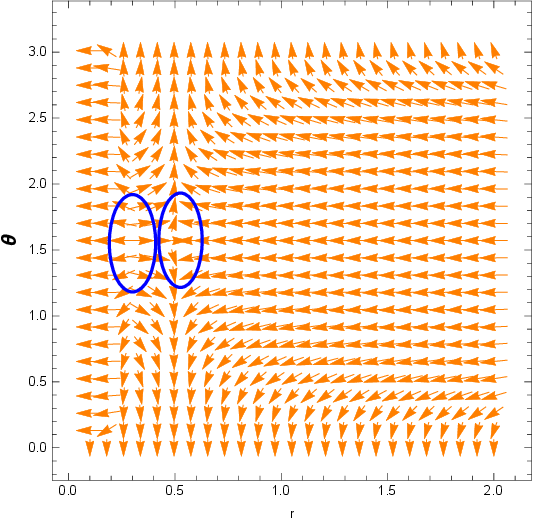}
 \label{1d}}\\
 \subfigure[]{
 \includegraphics[height=4cm,width=4cm]{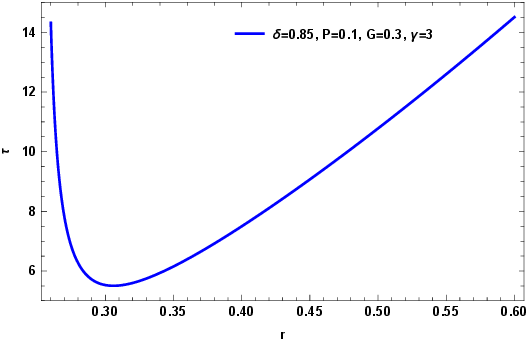}
 \label{1e}}
 \subfigure[]{
 \includegraphics[height=4cm,width=4cm]{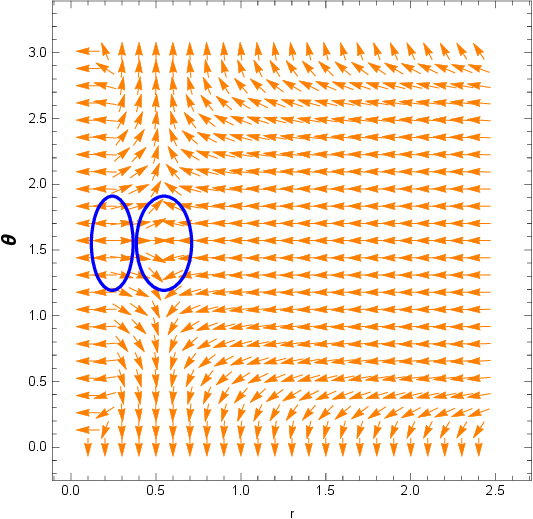}
 \label{1f}}
 \subfigure[]{
 \includegraphics[height=4cm,width=4cm]{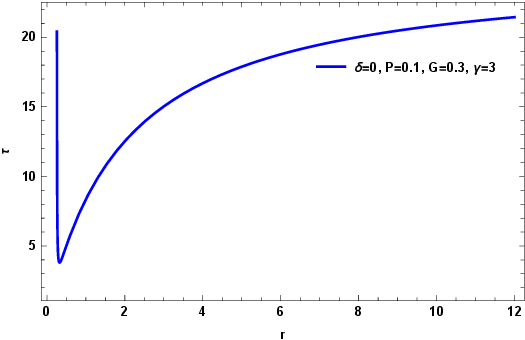}
 \label{1g}}
 \subfigure[]{
 \includegraphics[height=4cm,width=4cm]{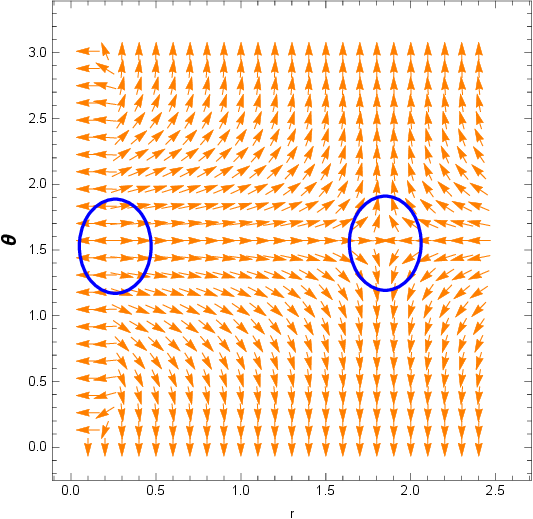}
 \label{1h}}
  \caption{\small{The curve represented by Equation (\ref{BB4}) is illustrated in Figures (\ref{1a}), (\ref{1c}), (\ref{1e}), and (\ref{1g}). In Figures (\ref{1b}), (\ref{1d}), (\ref{1f}), and (\ref{1h}), the zero points (ZPs) are positioned at coordinates $(r, \theta)$ with nonextensive parameter $\delta$.}}
 \label{m1}
 \end{center}
 \end{figure}

Now, we study the thermodynamic topology for the AdS Einstein-Power-Yang-Mills black holes from the one of the non extensive entropy viz Barrow within bulk boundary thermodynamics. So, With respect to above concepts and Eqs. (\ref{N2}), (\ref{F1}), and (\ref{BB3}), we can calculate the t$\mathcal{F}$.
\begin{equation}\label{BB1}
\mathcal{F}=\frac{r}{2 G} \left(\frac{8}{3} \pi  G P r+\frac{2^{\gamma -1} G \left(q^2\right)^{\gamma } r^{2-4 \gamma }}{4 \gamma -3}-\frac{2 \pi ^{\frac{\delta }{2}+1} r \left(\frac{r^2}{G}\right)^{\delta /2}}{\tau }+1\right)
\end{equation}
To study the black hole's topological charge, we need to obtain two vectors $\phi^{r_h}$ and $\phi^{\theta }$ using Eqs. (\ref{F2}),
\begin{equation}\label{BB2}
\phi^r=\frac{1}{12} \left(\frac{6 \left(\tau -2 \pi ^{\frac{\delta }{2}+1} (\delta +2) r \left(\frac{r^2}{G}\right)^{\delta /2}\right)}{G \tau }+32 \pi  P r-3\ 2^{\gamma } \left(q^2\right)^{\gamma } r^{2-4 \gamma }\right)
\end{equation}
and
\begin{equation}\label{BB3}
\phi^{\theta }=-\frac{\cot (\theta )}{\sin (\theta )}
\end{equation}
Here, the $\tau$ is calculated as,
\begin{equation}\label{BB4}
\tau =\frac{12 \pi ^{\frac{\delta }{2}+1} (\delta +2) r^{4 \gamma +1} \left(\frac{r^2}{G}\right)^{\delta /2}}{32 \pi  G P r^{4 \gamma +1}-3\ 2^{\gamma } G r^2 \left(q^2\right)^{\gamma }+6 r^{4 \gamma }}
\end{equation}
In our research, we delve into the thermodynamic topology of AdS Einstein-power-Yang-Mills black holes, examining them through the bulk-boundary and RPS frameworks. We consider various non-extensive entropy models, including Barrow, Rényi, Sharma-Mittal, Kaniadakis, and Tsallis-Cirto entropy. Initially, we analyze the thermodynamic topology within the bulk-boundary framework. The results are illustrated with normalized field lines on the right side of the figures. Figures (\ref{m1}) to (\ref{m5}) present the outcomes for Barrow, Rényi, Sharma-Mittal, Kaniadakis, and Tsallis-Cirto entropy, respectively. Figures (\ref{1b}), (\ref{1d}), (\ref{1f}), and (\ref{1h}) show two zero points, indicating topological charges influenced by the free parameters and the non-extensive parameter $\delta$. These charges, related to the winding number, are found within the blue contour loops at coordinates $(r, \theta)$. The order of these illustrations is determined by the parameter $\delta$.

Our findings reveal a unique characteristic: two topological charges $(\omega = +1, -1)$ and a total topological charge $W = 0$, represented by the zero points within the contour.

Moreover, as depicted in Figure (\ref{1h}), when the parameter $\delta$ is zero, our equations simplify to the Bekenstein-Hawking entropy structure, producing the same results as Barrow entropy.
\subsubsection{Rényi entropy, bulk boundary thermodynamics and thermodynamic topology}
\begin{figure}[h!]
 \begin{center}
 \subfigure[]{
 \includegraphics[height=4cm,width=4cm]{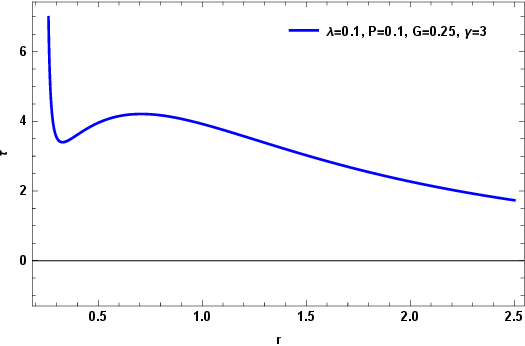}
 \label{2a}}
 \subfigure[]{
 \includegraphics[height=4cm,width=4cm]{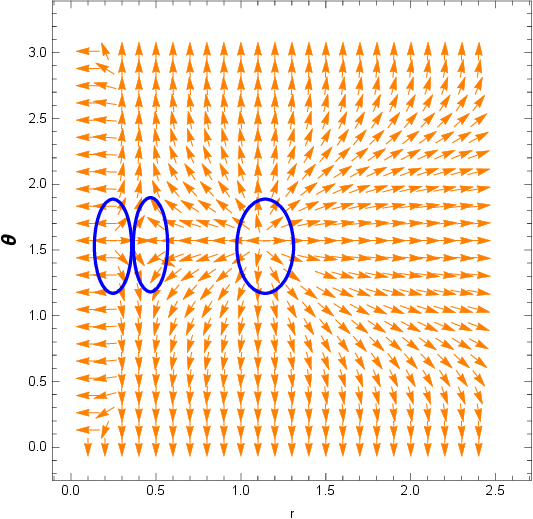}
 \label{2b}}
 \subfigure[]{
 \includegraphics[height=4cm,width=4cm]{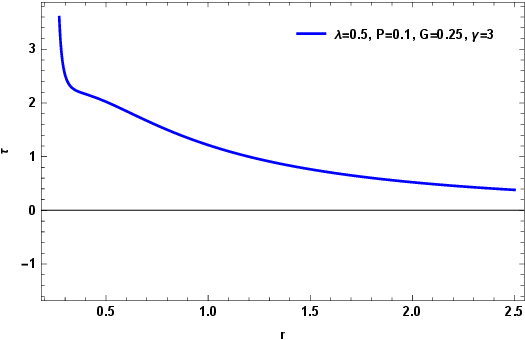}
 \label{2c}}
 \subfigure[]{
 \includegraphics[height=4cm,width=4cm]{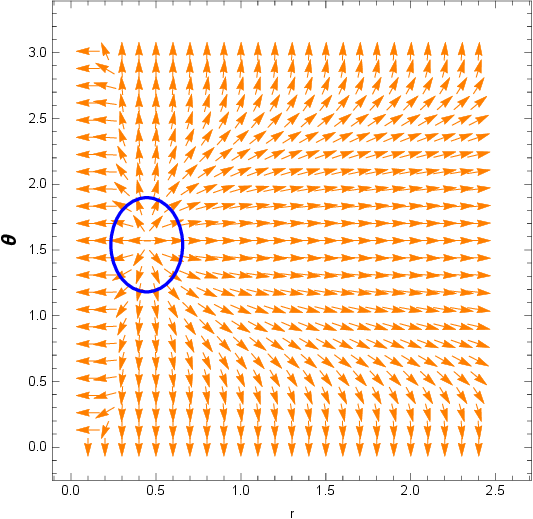}
 \label{2d}}\\
 \subfigure[]{
 \includegraphics[height=4cm,width=4cm]{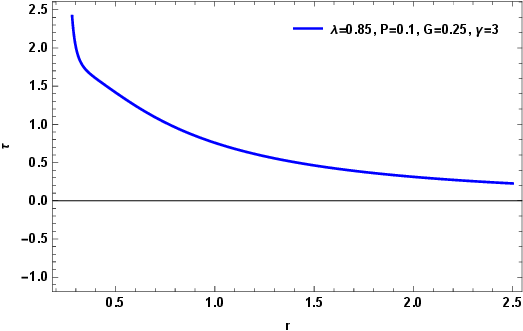}
 \label{2e}}
 \subfigure[]{
 \includegraphics[height=4cm,width=4cm]{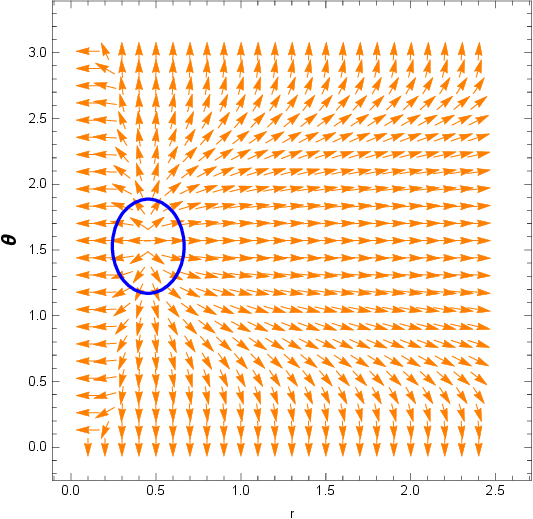}
 \label{2f}}
 \subfigure[]{
 \includegraphics[height=4cm,width=4cm]{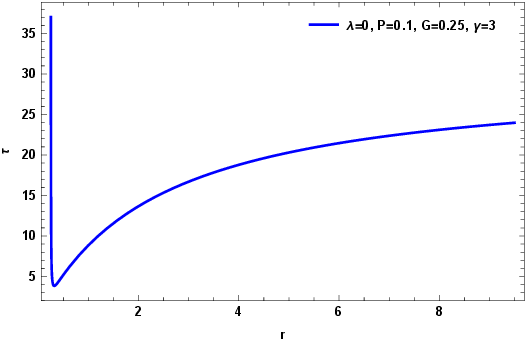}
 \label{2g}}
 \subfigure[]{
 \includegraphics[height=4cm,width=4cm]{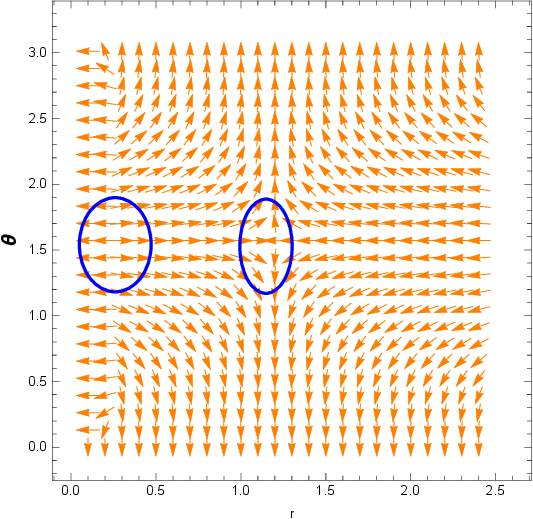}
 \label{2h}}
  \caption{\small{The curve represented by Equation (\ref{BR3}) is shown in Figures (\ref{2a}), (\ref{2c}), (\ref{2e}), and (\ref{2g}). In Figures (\ref{2b}), (\ref{2d}), (\ref{2f}), and (\ref{2h}), the zero points (ZPs) are positioned at coordinates $(r, \theta)$ with nonextensive parameter $\lambda$.}}
 \label{m2}
 \end{center}
 \end{figure}

We continue our investigation with Rényi entropy for the mentioned model in bulk boundary thermodynamics. So by using  Eqs. (\ref{N3}), (\ref{F1}), and (\ref{BB3}), the $\mathcal{F}$ is calculated as,
\begin{equation}\label{BR1}
\begin{split}
\mathcal{F}=\frac{1}{12} r \left(\frac{6}{G}+16 \pi  P r+\frac{3\ 2^{\gamma } \left(q^2\right)^{\gamma } r^{2-4 \gamma }}{4 \gamma -3}\right)-\frac{\log \left(\frac{\pi  \lambda  r^2}{G}+1\right)}{\lambda  \tau }
\end{split}
\end{equation}
Then, we obtain the $\phi^{r_h}$ with Eqs. (\ref{F2}) as follows,
\begin{equation}\label{BR2}
\begin{split}
\phi^r=-\frac{2 \pi  r}{G \tau +\pi  \lambda  r^2 \tau }+\frac{1}{2 G}+\frac{8 \pi  P r}{3}-2^{\gamma -2} \left(q^2\right)^{\gamma } r^{2-4 \gamma }
\end{split}
\end{equation}
Also, we can calculate $\tau$ with respect to (\ref{BR2}),
\begin{equation}\label{BR3}
\tau =\frac{24 \pi  G r^{4 \gamma +1}}{\left(G+\pi  \lambda  r^2\right) \left(32 \pi  G P r^{4 \gamma +1}-3\ 2^{\gamma } G r^2 \left(q^2\right)^{\gamma }+6 r^{4 \gamma }\right)}
\end{equation}
Figure (\ref{m2}) illustrates the results for Rényi entropy. Different topological charges are observed depending on the value of $\lambda$. When $\lambda = 0.1$, there are three topological charges $(\omega = +1, -1, +1)$ with a total topological charge $W = +1$. As $\lambda$ increases, the number of topological charges reduces to $(\omega = +1)$, maintaining a total topological charge $W = +1$. Setting $\lambda$ to zero results in two topological charges $(\omega = +1, -1)$ with a total topological charge $W = 0$. The presence of the Rényi non-extensive parameter alters the total topological charges compared to the Bekenstein-Hawking entropy.
\subsubsection{Sharma-Mittal entropy, bulk boundary thermodynamics and thermodynamic topology}
\begin{figure}[h!]
 \begin{center}
 \subfigure[]{
 \includegraphics[height=4cm,width=4cm]{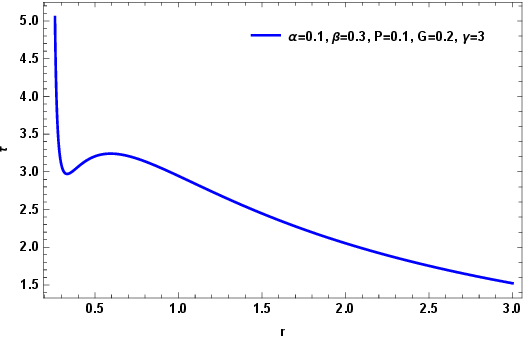}
 \label{3a}}
 \subfigure[]{
 \includegraphics[height=4cm,width=4cm]{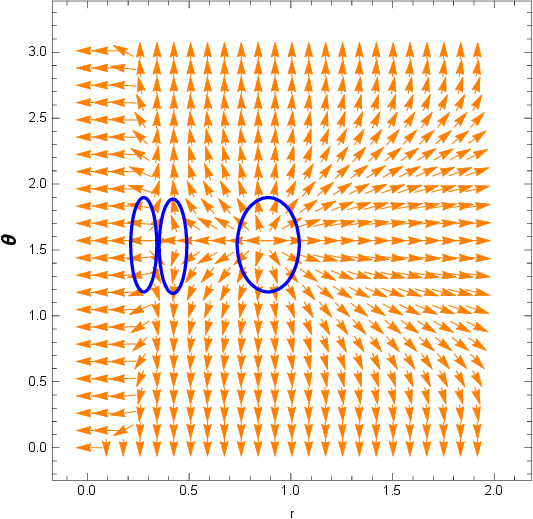}
 \label{3b}}
 \subfigure[]{
 \includegraphics[height=4cm,width=4cm]{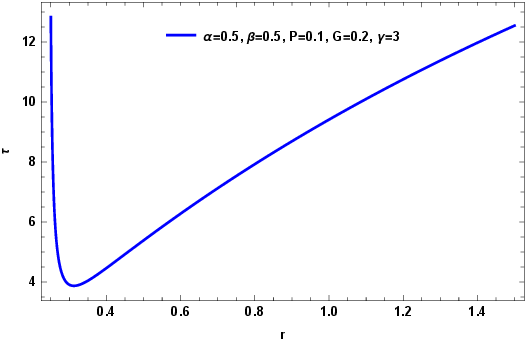}
 \label{3c}}
 \subfigure[]{
 \includegraphics[height=4cm,width=4cm]{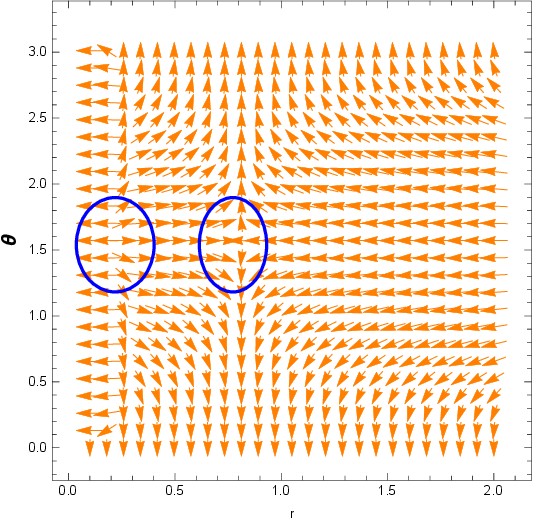}
 \label{3d}}\\
 \subfigure[]{
 \includegraphics[height=4cm,width=4cm]{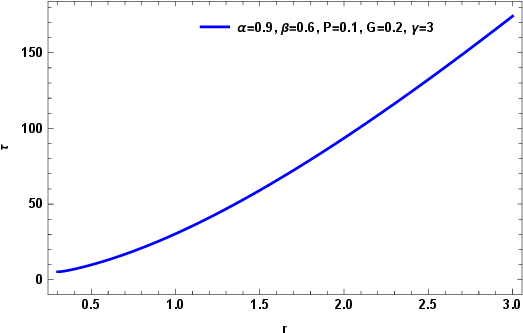}
 \label{3e}}
 \subfigure[]{
 \includegraphics[height=4cm,width=4cm]{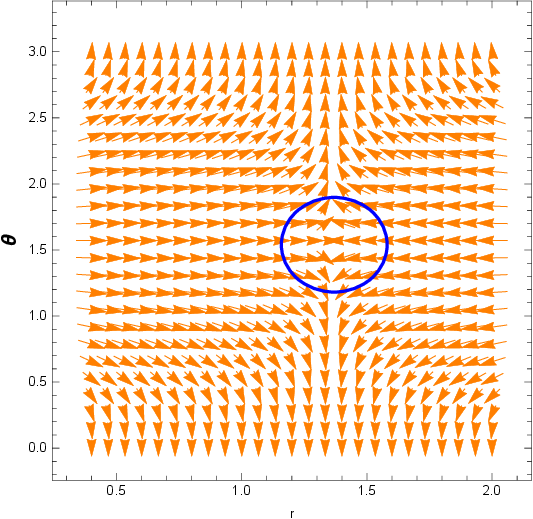}
 \label{3f}}
  \caption{\small{The curve represented by Equation (\ref{BSM3}) is depicted in Figures (\ref{3a}), (\ref{3c}), and (\ref{3e}). In Figures (\ref{3b}), (\ref{3d}), and (\ref{3f}), the zero points (ZPs) are positioned at coordinates $(r, \theta)$ with nonextensive parameters $(\alpha)$ and $(\beta)$.}}
 \label{m3}
 \end{center}
 \end{figure}

Now, We carry out our calculations for Sharma-Mittal entropy in the bulk boundary thermodynamics. With respect to Eqs. (\ref{N4}), (\ref{F1}), and (\ref{bB3}), we will have,
\begin{equation}\label{BSM1}
\mathcal{F}=\frac{1}{12} r \left(\frac{6}{G}+16 \pi  P r+\frac{3\ 2^{\gamma } \left(q^2\right)^{\gamma } r^{2-4 \gamma }}{4 \gamma -3}\right)-\frac{\left(\frac{\pi  \beta  r^2}{G}+1\right)^{\alpha /\beta }-1}{\alpha  \tau }
\end{equation}
Here, the $\phi^{r_h}$ is as follows,
\begin{equation}\label{BSM2}
\phi^r=\frac{\frac{1}{2}-\frac{2 \pi  r \left(\frac{\pi  \beta  r^2}{G}+1\right)^{\frac{\alpha }{\beta }-1}}{\tau }}{G}+\frac{8 \pi  P r}{3}-2^{\gamma -2} \left(q^2\right)^{\gamma } r^{2-4 \gamma }
\end{equation}
The $\tau$ is obtained as,
\begin{equation}\label{BSM3}
\tau =\frac{24 \pi  G r^{4 \gamma +1} \left(\frac{\pi  \beta  r^2}{G}+1\right)^{\alpha /\beta }}{\left(G+\pi  \beta  r^2\right) \left(32 \pi  G P r^{4 \gamma +1}-3\ 2^{\gamma } G r^2 \left(q^2\right)^{\gamma }+6 r^{4 \gamma }\right)}
\end{equation}
As shown in Figure (\ref{m3}), we observe different topological charges across three ranges for Sharma-Mittal entropy. When $\beta$ exceeds $\alpha$, there are three topological charges $(\omega = +1, -1, +1)$ with a total topological charge $W = +1$. When $\beta = \alpha$, two topological charges $(\omega = +1, -1)$ result in a total topological charge $W = 0$. When $\alpha$ exceeds $\beta$, a single topological charge $(\omega = -1)$ leads to a total topological charge $W = -1$, as depicted in Figures (\ref{3b}), (\ref{3d}), and (\ref{3f}).
\subsubsection{Kaniadakis entropy, bulk boundary thermodynamics and thermodynamic topology}
\begin{figure}[h!]
 \begin{center}
 \subfigure[]{
 \includegraphics[height=4cm,width=4cm]{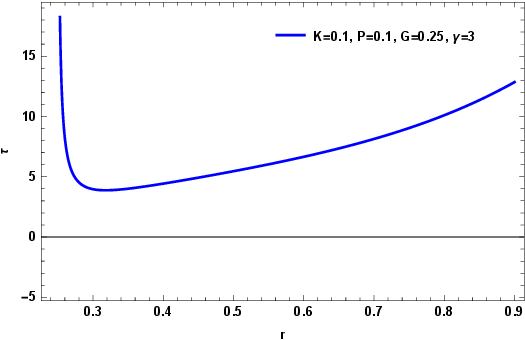}
 \label{4a}}
 \subfigure[]{
 \includegraphics[height=4cm,width=4cm]{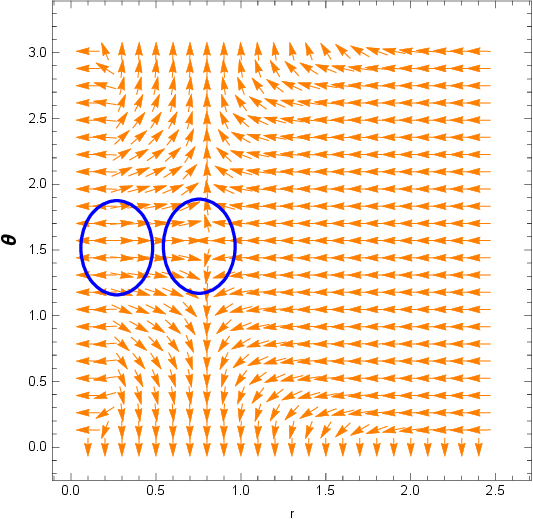}
 \label{4b}}
 \subfigure[]{
 \includegraphics[height=4cm,width=4cm]{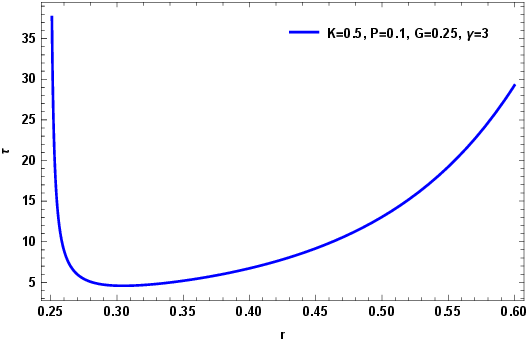}
 \label{4c}}
 \subfigure[]{
 \includegraphics[height=4cm,width=4cm]{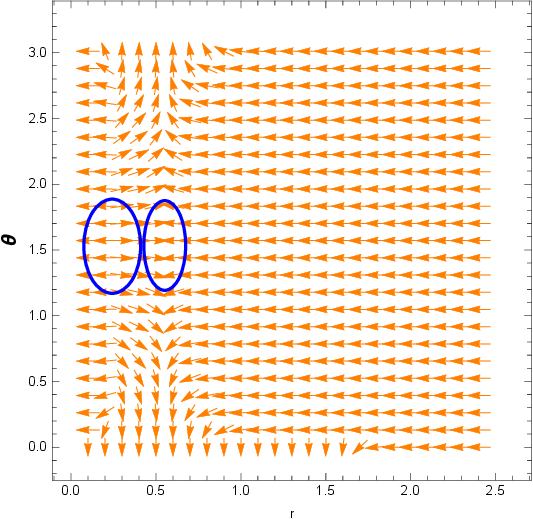}
 \label{4d}}\\
 \subfigure[]{
 \includegraphics[height=4cm,width=4cm]{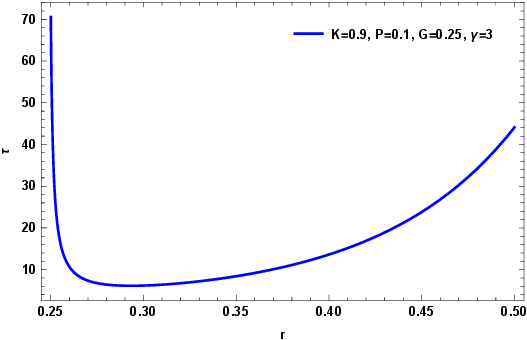}
 \label{4e}}
 \subfigure[]{
 \includegraphics[height=4cm,width=4cm]{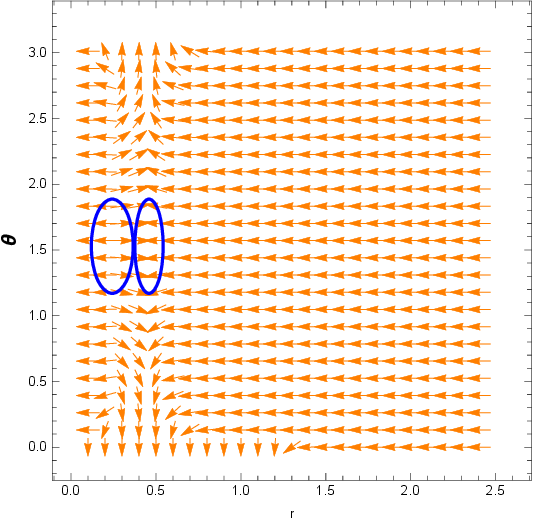}
 \label{4f}}
 \subfigure[]{
 \includegraphics[height=4cm,width=4cm]{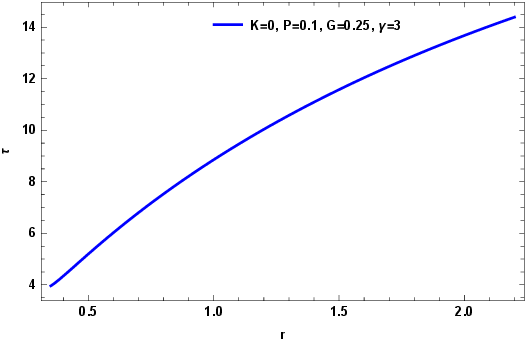}
 \label{4g}}
 \subfigure[]{
 \includegraphics[height=4cm,width=4cm]{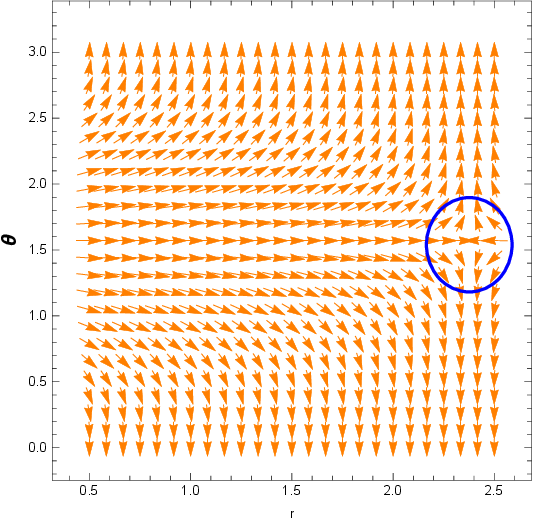}
 \label{4h}}
  \caption{\small{The curve represented by Equation (\ref{BK3}) is shown in Figures (\ref{4a}), (\ref{4c}), (\ref{4e}), and (\ref{4g}). In Figures (\ref{4b}), (\ref{4d}), (\ref{4f}), and (\ref{4h}), the zero points (ZPs) are positioned at coordinates $(r, \theta)$ with nonextensive parameter $K$.}}
 \label{m4}
 \end{center}
 \end{figure}

Here, for Kaniadakis entropy and using the Eqs. (\ref{N5}), (\ref{F1}), and (\ref{BB3}), one can obtain,
\begin{equation}\label{BK1}
\mathcal{F}=\frac{1}{12} r \left(\frac{6}{G}+16 \pi  P r+\frac{3\ 2^{\gamma } \left(q^2\right)^{\gamma } r^{2-4 \gamma }}{4 \gamma -3}\right)-\frac{\sinh \left(\frac{\pi  K r^2}{G}\right)}{K \tau }
\end{equation}
Now, the $\phi^{r_h}$ is determined,
\begin{equation}\label{BK2}
\phi^r=\frac{1}{12} \left(\frac{6}{G}+32 \pi  P r-3\ 2^{\gamma } \left(q^2\right)^{\gamma } r^{2-4 \gamma }\right)-\frac{2 \pi  r \cosh \left(\frac{\pi  K r^2}{G}\right)}{G \tau }
\end{equation}
We obtain $\tau$ as,
\begin{equation}\label{BK3}
\tau =\frac{24 \pi  r^{4 \gamma +1} \cosh \left(\frac{\pi  K r^2}{G}\right)}{32 \pi  G P r^{4 \gamma +1}-3\ 2^{\gamma } G r^2 \left(q^2\right)^{\gamma }+6 r^{4 \gamma }}
\end{equation}
Figure (\ref{m4}) shows Kaniadakis entropy, where the number of topological charges is influenced by the Sharma-Mittal entropy non-extensive parameter $K$. We observe two topological charges $(\omega = +1, -1)$ with a total topological charge $W = 0$ for any acceptable value of $K$, except when $K = 0$, where a single topological charge $(\omega = -1)$ results in a total topological charge $W = -1$, as shown in Figure (\ref{4h}).
\subsubsection{Tsallis-Cirto entropy, bulk boundary thermodynamics and thermodynamic topology}
\begin{figure}[h!]
 \begin{center}
 \subfigure[]{
 \includegraphics[height=4cm,width=4cm]{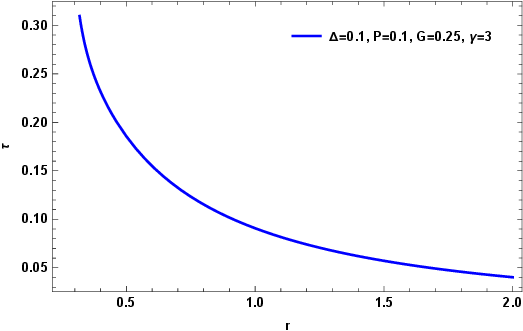}
 \label{5a}}
 \subfigure[]{
 \includegraphics[height=4cm,width=4cm]{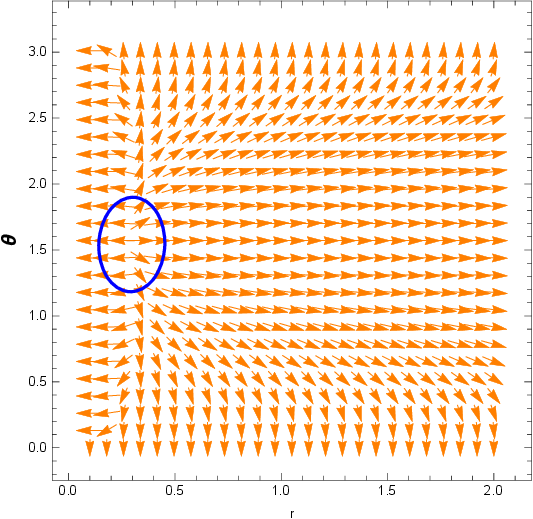}
 \label{5b}}
 \subfigure[]{
 \includegraphics[height=4cm,width=4cm]{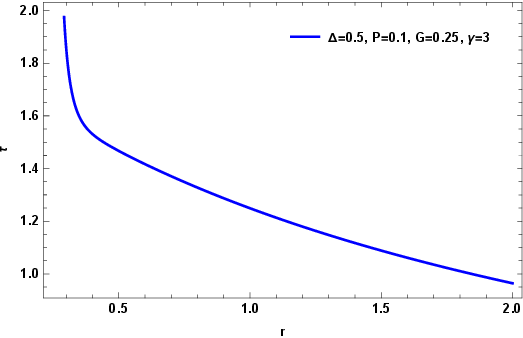}
 \label{5c}}
 \subfigure[]{
 \includegraphics[height=4cm,width=4cm]{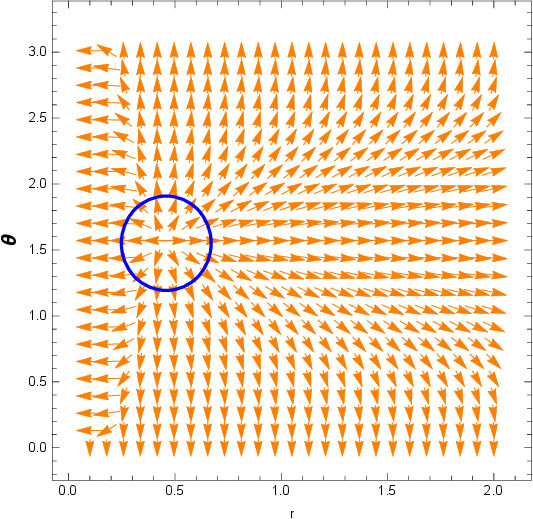}
 \label{5d}}\\
 \subfigure[]{
 \includegraphics[height=4cm,width=4cm]{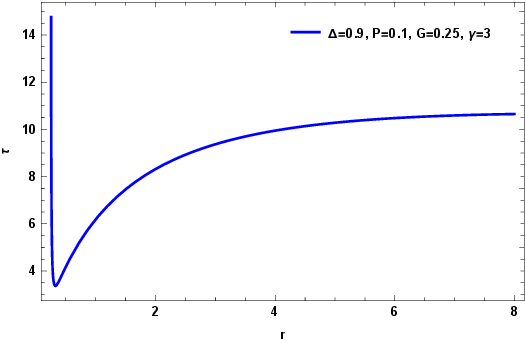}
 \label{5e}}
 \subfigure[]{
 \includegraphics[height=4cm,width=4cm]{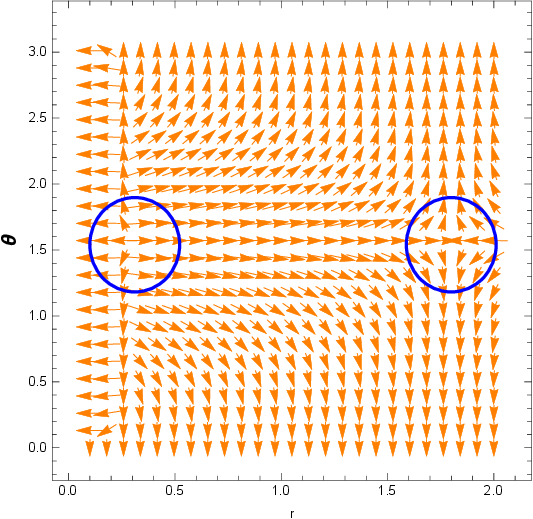}
 \label{5f}}
  \caption{\small{The curve represented by Equation (\ref{BTC3}) is depicted in Figures (\ref{5a}), (\ref{5c}), and (\ref{5e}). In Figures (\ref{5b}), (\ref{5d}), and (\ref{5f}), the zero points (ZPs) are positioned at coordinates $(r, \theta)$ on the circular loops with nonextensive parameter $\Delta$.}}
 \label{m5}
 \end{center}
 \end{figure}

Also, like the pervious subsection, we study thermodymaic topology for Tsallis-Cirto entropy in the bulk boundary thermodynamics. So by using the Eqs. (\ref{N6}), (\ref{F1}), and (\ref{BB3}), we can calculate
\begin{equation}\label{BTC1}
\mathcal{F}=-\frac{\pi ^{\Delta } \left(\frac{r^2}{G}\right)^{\Delta }}{\tau }+\frac{r}{2 G}+\frac{4}{3} \pi  P r^2+\frac{2^{\gamma -2} \left(q^2\right)^{\gamma } r^{3-4 \gamma }}{4 \gamma -3}
\end{equation}
We obtain the $\phi^{r_h}$,
\begin{equation}\label{BTC2}
\phi^r=-\frac{2 \pi ^{\Delta } \Delta  \left(\frac{r^2}{G}\right)^{\Delta }}{r \tau }+\frac{1}{2 G}+\frac{8 \pi  P r}{3}-2^{\gamma -2} \left(q^2\right)^{\gamma } r^{2-4 \gamma }
\end{equation}
The $\tau$ is calculated as,
\begin{equation}\label{BTC3}
\tau =\frac{24 \pi ^{\Delta } \Delta  G r^{4 \gamma -1} \left(\frac{r^2}{G}\right)^{\Delta }}{32 \pi  G P r^{4 \gamma +1}-3\ 2^{\gamma } G r^2 \left(q^2\right)^{\gamma }+6 r^{4 \gamma }}
\end{equation}
Figure (\ref{m5}) presents Tsallis-Cirto entropy. For small parameter $\Delta$ values, a single topological charge $(\omega = +1)$ results in a total topological charge $W = +1$. When $\Delta$ increases to 0.9, two topological charges $(\omega = +1, -1)$ appear, with a total topological charge $W = 0$. This indicates that the non-extensive parameters significantly affect the classification and number of topological charges, underscoring the importance of these non-extensive entropies compared to the traditional Bekenstein-Hawking case.
\subsection{Restricted phase space thermodynamics}
Here, we will derive the equations for AdS EPYM black holes in RPS thermodynamics. The entropy $S$ for this black holes is defined as,
\begin{equation}\label{RPS1}
q = \frac{\hat{q}}{\sqrt{C}}, \quad G = \frac{l^2}{C}, \quad S = \frac{C r_h^2 \pi}{l^2}
\end{equation}
Thus, the temperature $T$ of AdS EPYM black holes in RPS thermodynamics is given by:
\begin{equation}\label{RPS2}
T = \frac{1 + \frac{r_h^2}{l^2} - \left(\frac{2\hat{q}^2}{C}\right)^\gamma \frac{l^2}{2C r_h^{4\gamma - 2}}}{4\pi r_h}
\end{equation}
The parameter $C$, based on the provided equations, is calculated as,
\begin{equation}\label{RPS3}
C = 2\hat{q}^2 \exp\left(-\frac{2 \ln(2) - \ln(4\gamma - 1) + \ln\left(\frac{\hat{q}^2 r_h^{4\gamma - 2} (l^2 - r_h^2)}{l^4}\right)}{\gamma + 1}\right)
\end{equation}
We can determine the mass $M$ for this black hole as follows,
\begin{equation}\label{RPS4}
M = \frac{r_h \left(1 + \frac{r_h^2}{l^2} - \left(\frac{\hat{q}}{\sqrt{C}}\right)^{2\gamma} \frac{2^{-1+\gamma} l^2}{C r_h^{4\gamma - 2} (4\gamma - 3)}\right) C}{2l^2}.
\end{equation}
\subsubsection{Barrow entropy, RPS thermodynamics and thermodynamic topology}
In a manner similar to the previous section, we can explore the thermodynamic topology of AdS EPYM black holes with respect to non-extensive entropy within the context of RPS thermodynamics. For Barrow entropy, by utilizing Equations (\ref{N2}), (\ref{F1}), and (\ref{RPS4}), we can calculate the $\mathcal{F}$.
\begin{equation}\label{RB1}
\mathcal{F}=\frac{1}{4} r \left(\frac{2 C \left(l^2 \left(1-\frac{2 \pi ^{\frac{\delta }{2}+1} r \left(\frac{C r^2}{l^2}\right)^{\delta /2}}{\tau }\right)+r^2\right)}{l^4}+\frac{2^{\gamma } r^{2-4 \gamma } \left(\frac{\hat{q}}{\sqrt{C}}\right)^{2 \gamma }}{4 \gamma -3}\right)
\end{equation}
Next, we compute $\phi^{r_h}$ using Equation (\ref{F2}) as follows:
\begin{equation}\label{RB2}
\phi^r=\frac{1}{4} \left(\frac{2 C \left(\frac{l^2 \left(\tau -2 \pi ^{\frac{\delta }{2}+1} (\delta +2) r \left(\frac{C r^2}{l^2}\right)^{\delta /2}\right)}{\tau }+3 r^2\right)}{l^4}-2^{\gamma } r^{2-4 \gamma } \left(\frac{\hat{q}}{\sqrt{C}}\right)^{2 \gamma }\right)
\end{equation}
We calculate the $(\tau)$ as follows,
\begin{equation}\label{RB3}
\tau =\frac{4 C \pi ^{\frac{\delta }{2}+1} (\delta +2) l^2 r^{4 \gamma +1} \left(\frac{C r^2}{l^2}\right)^{\delta /2}}{-2^{\gamma } l^4 r^2 \left(\frac{\hat{q}}{\sqrt{C}}\right)^{2 \gamma }+2 C l^2 r^{4 \gamma }+6 C r^{4 \gamma +2}}
\end{equation}

\begin{figure}[h!]
 \begin{center}
 \subfigure[]{
 \includegraphics[height=4cm,width=4cm]{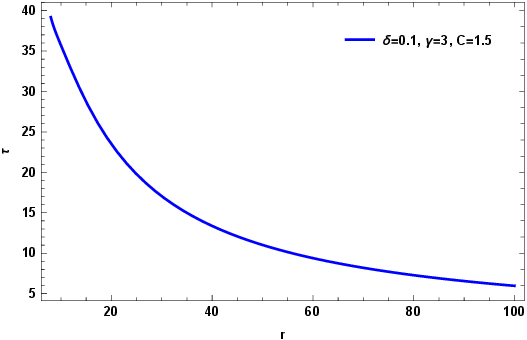}
 \label{6a}}
 \subfigure[]{
 \includegraphics[height=4cm,width=4cm]{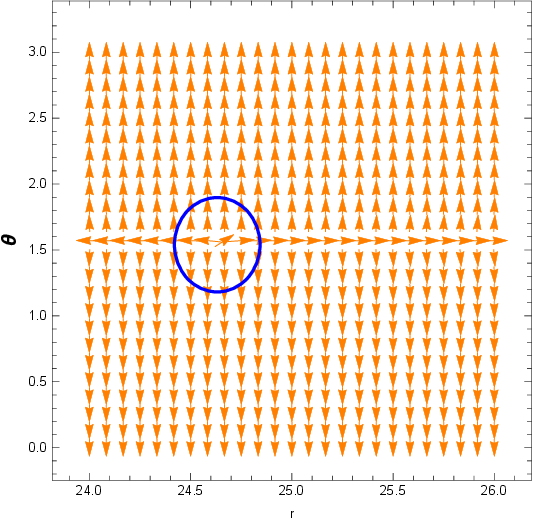}
 \label{6b}}
 \subfigure[]{
 \includegraphics[height=4cm,width=4cm]{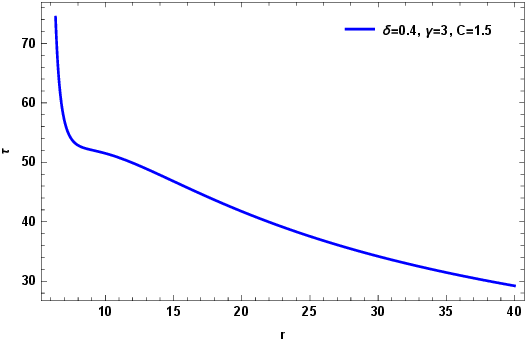}
 \label{6c}}
 \subfigure[]{
 \includegraphics[height=4cm,width=4cm]{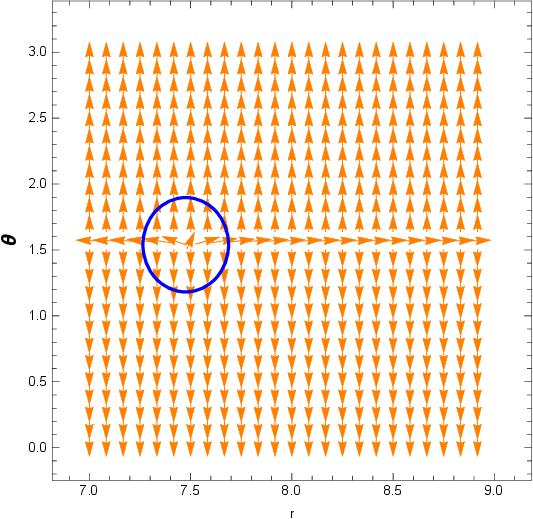}
 \label{6d}}
 \subfigure[]{
 \includegraphics[height=4cm,width=4cm]{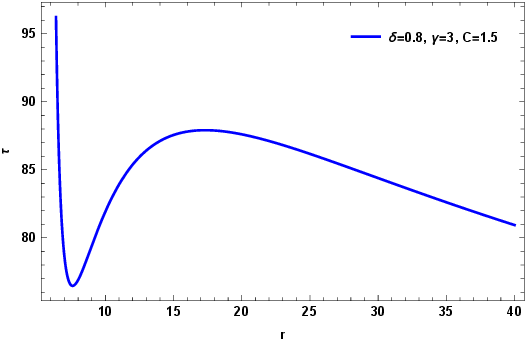}
 \label{6e}}
 \subfigure[]{
 \includegraphics[height=4cm,width=4cm]{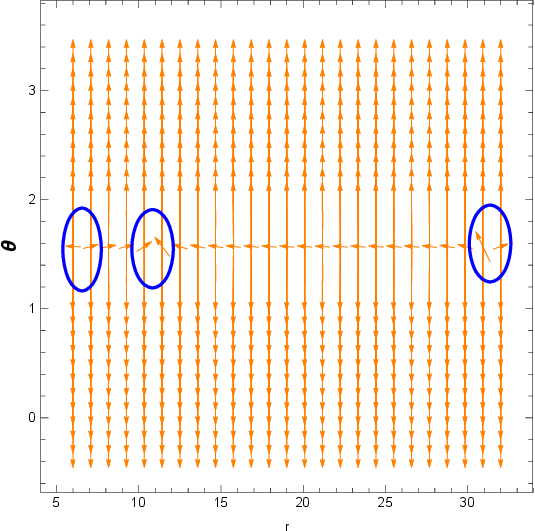}
 \label{6f}}
 \subfigure[]{
 \includegraphics[height=4cm,width=4cm]{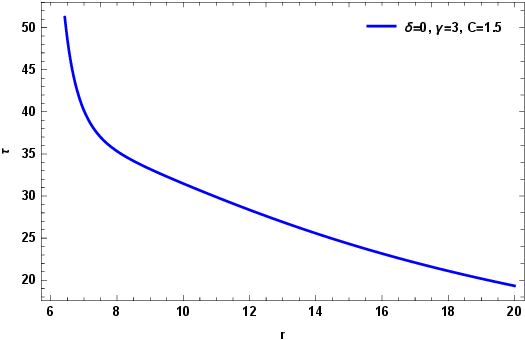}
 \label{6g}}
 \subfigure[]{
 \includegraphics[height=4cm,width=4cm]{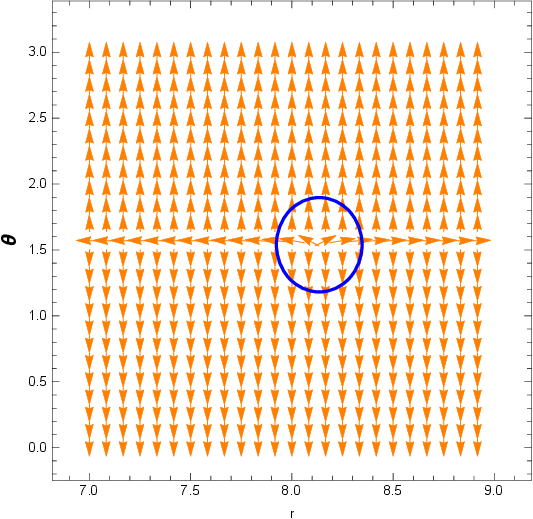}
 \label{6h}}
  \caption{\small{The curve represented by Equation (\ref{RB3}) is shown in Figures (\ref{6a}), (\ref{6c}), (\ref{6e}), and (\ref{6g}). In Figures (\ref{6b}), (\ref{6d}), (\ref{6f}), and (\ref{6h}), the zero points (ZPs) are positioned at coordinates $(r, \theta)$ with nonextensive parameter $\delta$.}}
 \label{m6}
 \end{center}
 \end{figure}
\subsubsection{Rényi entropy, RPS thermodynamics and thermodynamic topology}
Like the previous subsection for Rényi entropy, with Eqs. (\ref{N3}), (\ref{F1}), and (\ref{RPS4}), we can obtain the $\mathcal{F}$.
\begin{equation}\label{RR1}
\mathcal{F}=\frac{1}{4} \left(-\frac{4 \log \left(\frac{\pi  C \lambda  r^2}{l^2}+1\right)}{\lambda  \tau }+\frac{2 C r \left(l^2+r^2\right)}{l^4}+\frac{2^{\gamma } r^{3-4 \gamma } \left(\frac{\hat{q}}{\sqrt{C}}\right)^{2 \gamma }}{4 \gamma -3}\right)
\end{equation}
Next, we compute $\phi^{r_h}$ using Equation (\ref{F2}) as follows:,
\begin{equation}\label{RR2}
\phi^r=\frac{1}{4} \left(2 C \left(-\frac{4 \pi  r}{\pi  C \lambda  r^2 \tau +l^2 \tau }+\frac{3 r^2}{l^4}+\frac{1}{l^2}\right)-2^{\gamma } r^{2-4 \gamma } \left(\frac{\hat{q}}{\sqrt{C}}\right)^{2 \gamma }\right)
\end{equation}
Here, we will have,
\begin{equation}\label{RR3}
\tau =\frac{8 \pi  C l^4 r^{4 \gamma +1}}{\left(\pi  C \lambda  r^2+l^2\right) \left(-2^{\gamma } l^4 r^2 \left(\frac{\hat{q}}{\sqrt{C}}\right)^{2 \gamma }+2 C l^2 r^{4 \gamma }+6 C r^{4 \gamma +2}\right)}
\end{equation}
\begin{figure}[h!]
 \begin{center}
 \subfigure[]{
 \includegraphics[height=4cm,width=4cm]{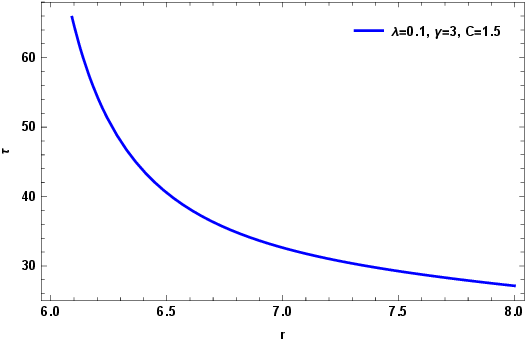}
 \label{7a}}
 \subfigure[]{
 \includegraphics[height=4cm,width=4cm]{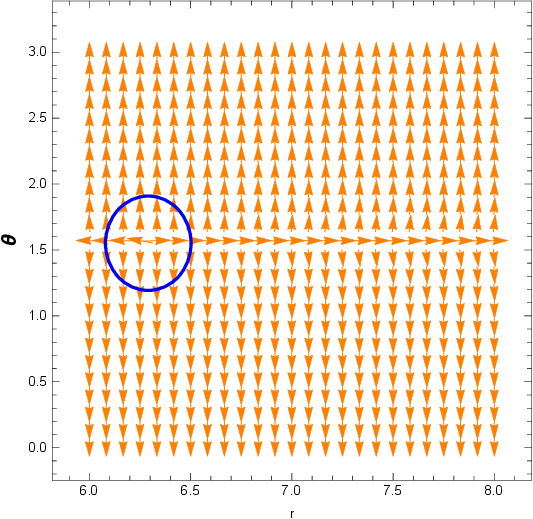}
 \label{7b}}
 \subfigure[]{
 \includegraphics[height=4cm,width=4cm]{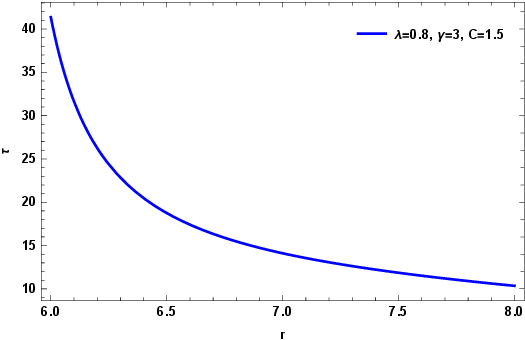}
 \label{7c}}
 \subfigure[]{
 \includegraphics[height=4cm,width=4cm]{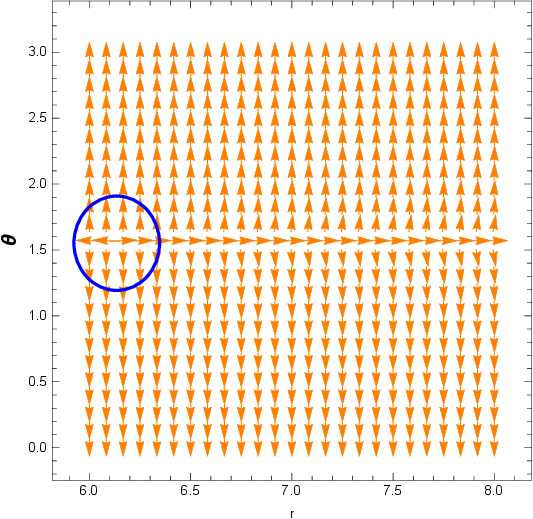}
 \label{7d}}
 \subfigure[]{
 \includegraphics[height=4cm,width=4cm]{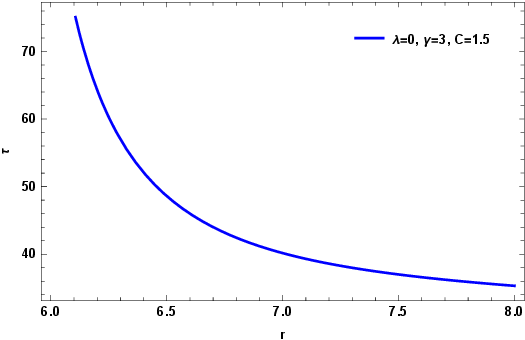}
 \label{7e}}
 \subfigure[]{
 \includegraphics[height=4cm,width=4cm]{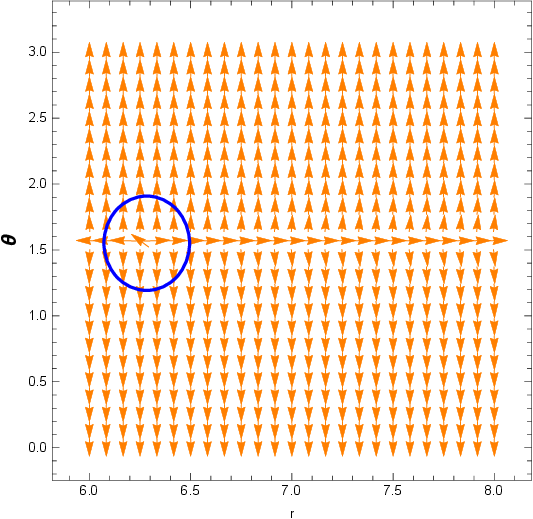}
 \label{7f}}
  \caption{\small{The curve represented by Equation (\ref{RR3}) is depicted in Figures (\ref{7a}), (\ref{7c}), and (\ref{7e}). In Figures (\ref{7b}), (\ref{7d}), and (\ref{7f}), the zero points (ZPs) are positioned at coordinates $(r, \theta)$, corresponding to the nonextensive parameter $(\lambda)$.}}
 \label{m7}
 \end{center}
 \end{figure}
\subsubsection{Sharma-Mittal entropy, RPS thermodynamics and thermodynamic topology}
Now, with respect to Eqs. (\ref{N4}), (\ref{F1}), and (\ref{RPS4}) we obtain,
\begin{equation}\label{RSM1}
\mathcal{F}=\frac{1}{4} \left(-\frac{4 \left(\left(\frac{\pi  \beta  C r^2}{l^2}+1\right)^{\alpha /\beta }-1\right)}{\alpha  \tau }+\frac{2 C r \left(l^2+r^2\right)}{l^4}+\frac{2^{\gamma } r^{3-4 \gamma } \left(\frac{\hat{q}}{\sqrt{C}}\right)^{2 \gamma }}{4 \gamma -3}\right)
\end{equation}
The $\phi^{r_h}$ is calculated as follows,
\begin{equation}\label{RSM2}
\phi^r=\frac{1}{4} \left(\frac{2 C \left(l^2 \left(1-\frac{4 \pi  r \left(\frac{\pi  \beta  C r^2}{l^2}+1\right)^{\frac{\alpha }{\beta }-1}}{\tau }\right)+3 r^2\right)}{l^4}-2^{\gamma } r^{2-4 \gamma } \left(\frac{\hat{q}}{\sqrt{C}}\right)^{2 \gamma }\right)
\end{equation}
Then we determine the $\tau$,
\begin{equation}\label{RSM3}
\tau =\frac{8 \pi  C l^4 r^{4 \gamma +1} \left(\frac{\pi  \beta  C r^2}{l^2}+1\right)^{\alpha /\beta }}{2 \pi  \beta  C^2 l^2 r^{4 \gamma +2}+6 \pi  \beta  C^2 r^{4 \gamma +4}-2^{\gamma } l^6 r^2 \left(\frac{\hat{q}}{\sqrt{C}}\right)^{2 \gamma }-\pi  \beta  2^{\gamma } C l^4 r^4 \left(\frac{\hat{q}}{\sqrt{C}}\right)^{2 \gamma }+2 C l^4 r^{4 \gamma }+6 C l^2 r^{4 \gamma +2}}
\end{equation}

\begin{figure}[h!]
 \begin{center}
 \subfigure[]{
 \includegraphics[height=4cm,width=4cm]{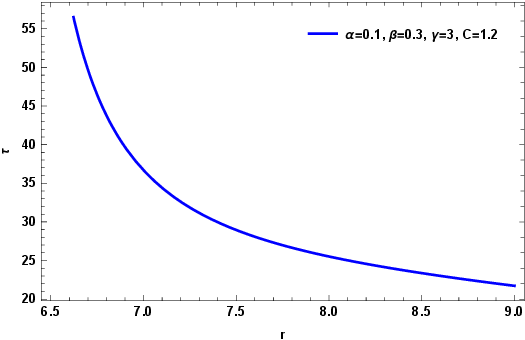}
 \label{8a}}
 \subfigure[]{
 \includegraphics[height=4cm,width=4cm]{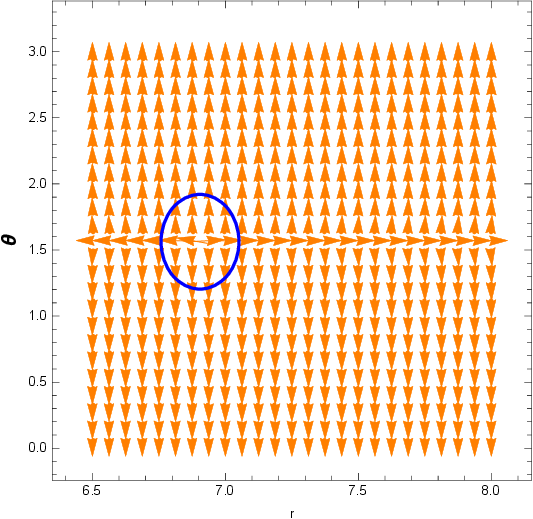}
 \label{8b}}
 \subfigure[]{
 \includegraphics[height=4cm,width=4cm]{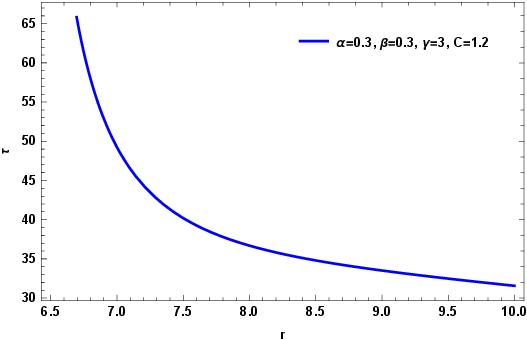}
 \label{8c}}
 \subfigure[]{
 \includegraphics[height=4cm,width=4cm]{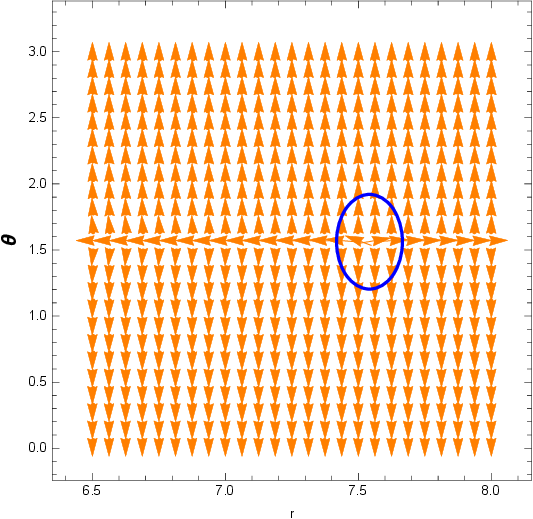}
 \label{8d}}
 \subfigure[]{
 \includegraphics[height=4cm,width=4cm]{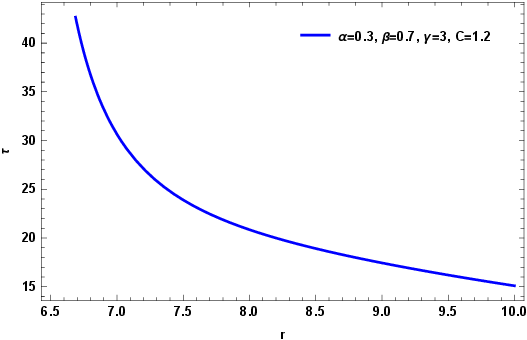}
 \label{8e}}
 \subfigure[]{
 \includegraphics[height=4cm,width=4cm]{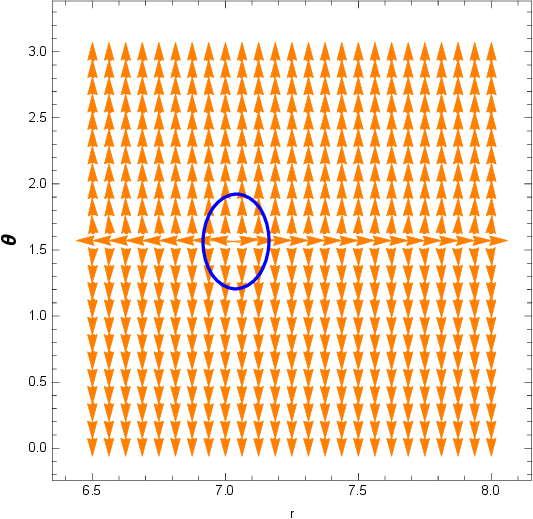}
 \label{8f}}
  \caption{\small{The curve represented by Equation (\ref{RSM3}) is illustrated in Figures (\ref{8a}), (\ref{8c}), and (\ref{8e}). In Figures (\ref{8b}), (\ref{8d}), and (\ref{8f}), the zero points (ZPs) are positioned at coordinates $(r, \theta)$, corresponding to the nonextensive parameters $(\alpha)$ and $(\beta)$.}}
 \label{m8}
 \end{center}
 \end{figure}
\subsubsection{Kaniadakis entropy, RPS thermodynamics and thermodynamic topology}
For Kaniadakis entropy in RPS thermodynamics and by using Eqs. (\ref{N5}), (\ref{F1}), and (\ref{RPS4}) we can calculate,
\begin{equation}\label{RK1}
\mathcal{F}=\frac{1}{4} \left(-\frac{4 \sinh \left(\frac{\pi  C K r^2}{l^2}\right)}{K \tau }+\frac{2 C r \left(l^2+r^2\right)}{l^4}+\frac{2^{\gamma } r^{3-4 \gamma } \left(\frac{\hat{q}}{\sqrt{C}}\right)^{2 \gamma }}{4 \gamma -3}\right)
\end{equation}
Then the $\phi^{r_h}$ is obtained as,
\begin{equation}\label{RK2}
\phi^r=\frac{1}{4} \left(-\frac{8 \pi  C r \cosh \left(\frac{\pi  C K r^2}{l^2}\right)}{l^2 \tau }+\frac{2 C \left(l^2+3 r^2\right)}{l^4}-2^{\gamma } r^{2-4 \gamma } \left(\frac{\hat{q}}{\sqrt{C}}\right)^{2 \gamma }\right)
\end{equation}
Also, we can calculate the $\tau$,
\begin{equation}\label{RK3}
\tau =\frac{8 \pi  C l^2 r^{4 \gamma +1} \cosh \left(\frac{\pi  C K r^2}{l^2}\right)}{-2^{\gamma } l^4 r^2 \left(\frac{\hat{q}}{\sqrt{C}}\right)^{2 \gamma }+2 C l^2 r^{4 \gamma }+6 C r^{4 \gamma +2}}
\end{equation}

\begin{figure}[h!]
 \begin{center}
 \subfigure[]{
 \includegraphics[height=4cm,width=4cm]{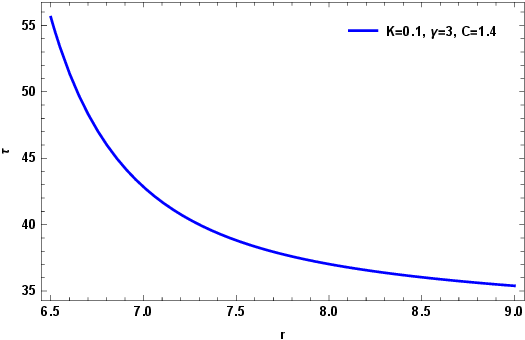}
 \label{9a}}
 \subfigure[]{
 \includegraphics[height=4cm,width=4cm]{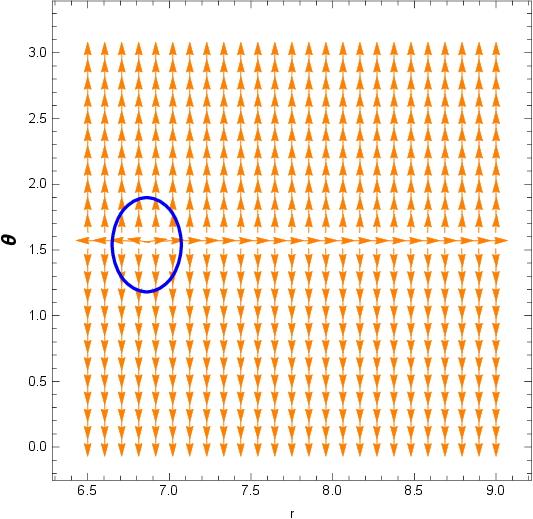}
 \label{9b}}
 \subfigure[]{
 \includegraphics[height=4cm,width=4cm]{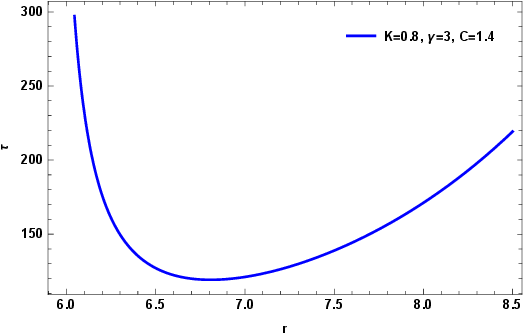}
 \label{9c}}
 \subfigure[]{
 \includegraphics[height=4cm,width=4cm]{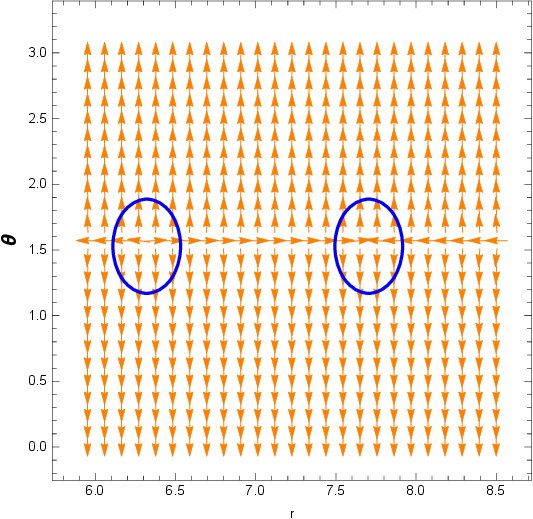}
 \label{9d}}
 \subfigure[]{
 \includegraphics[height=4cm,width=4cm]{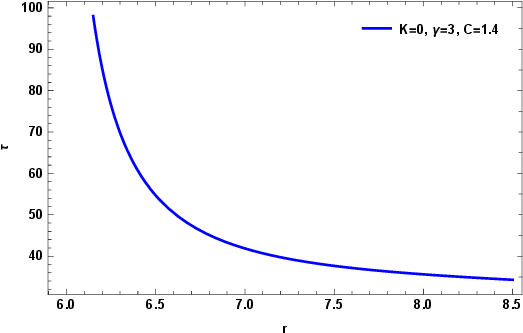}
 \label{9e}}
 \subfigure[]{
 \includegraphics[height=4cm,width=4cm]{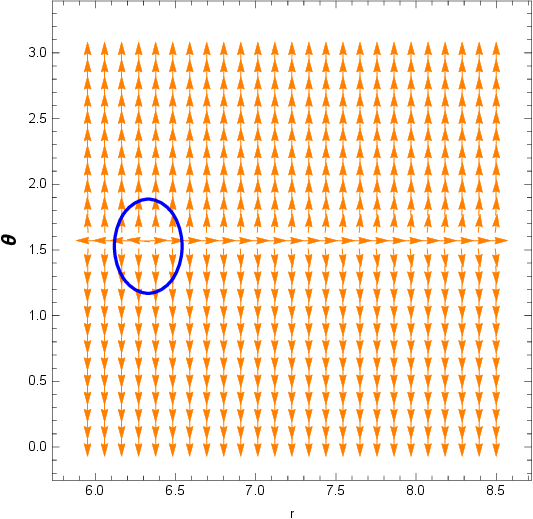}
 \label{9f}}
  \caption{\small{The curve represented by Equation (\ref{RK3}) is illustrated in Figures (\ref{9a}), (\ref{9c}), and (\ref{9e}). In Figures (\ref{9b}), (\ref{9d}), and (\ref{9f}), the zero points (ZPs) are positioned at coordinates $(r, \theta)$, corresponding to the nonextensive parameters $K$.}}
 \label{m9}
 \end{center}
 \end{figure}

\subsubsection{Tsallis-Cirto entropy, RPS thermodynamics and thermodynamic topology}
Here For Tsallis-Cirto in RPS thermodynamics for mentioned model and with Eqs. (\ref{N6}), (\ref{F1}), and (\ref{RPS4}) we will have,
\begin{equation}\label{RTC1}
\mathcal{F}=\frac{1}{4} \left(-\frac{4 \pi ^{\Delta } \left(\frac{C r^2}{l^2}\right)^{\Delta }}{\tau }+\frac{2 C r \left(l^2+r^2\right)}{l^4}+\frac{2^{\gamma } r^{3-4 \gamma } \left(\frac{\hat{q}}{\sqrt{C}}\right)^{2 \gamma }}{4 \gamma -3}\right)
\end{equation}
The $\phi^{r_h}$ is calculted as,
\begin{equation}\label{RTC2}
\phi^r=\frac{1}{4} \left(-\frac{8 \pi ^{\Delta } \Delta  \left(\frac{C r^2}{l^2}\right)^{\Delta }}{r \tau }+\frac{2 C \left(l^2+3 r^2\right)}{l^4}-2^{\gamma } r^{2-4 \gamma } \left(\frac{\hat{q}}{\sqrt{C}}\right)^{2 \gamma }\right)
\end{equation}
The $\tau$ is determined as,
\begin{equation}\label{RTC3}
\tau =\frac{8 \pi ^{\Delta } \Delta  l^4 r^{4 \gamma -1} \left(\frac{C r^2}{l^2}\right)^{\Delta }}{-2^{\gamma } l^4 r^2 \left(\frac{\hat{q}}{\sqrt{C}}\right)^{2 \gamma }+2 C l^2 r^{4 \gamma }+6 C r^{4 \gamma +2}}
\end{equation}

\begin{figure}[h!]
 \begin{center}
 \subfigure[]{
 \includegraphics[height=4cm,width=4cm]{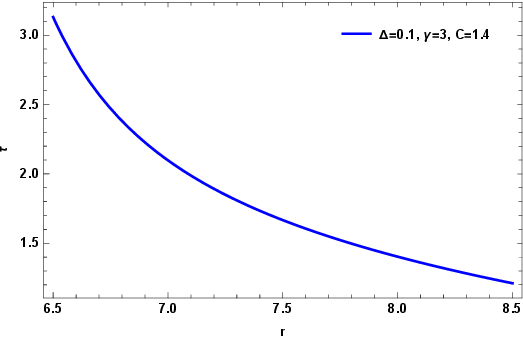}
 \label{10a}}
 \subfigure[]{
 \includegraphics[height=4cm,width=4cm]{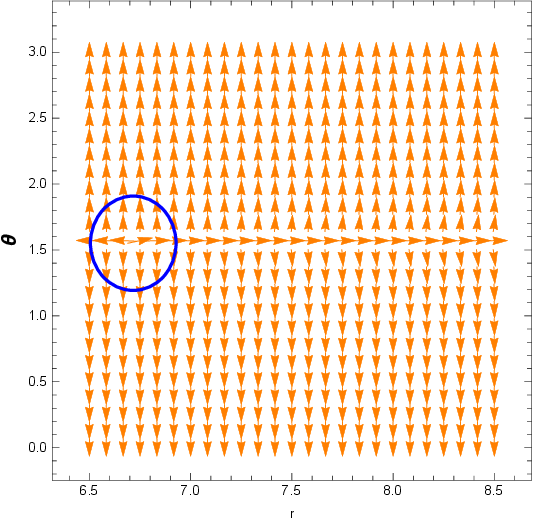}
 \label{10b}}
 \subfigure[]{
 \includegraphics[height=4cm,width=4cm]{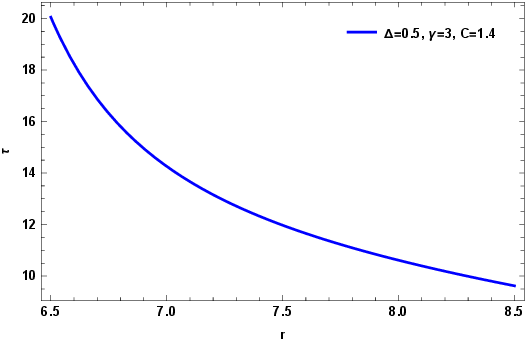}
 \label{10c}}
 \subfigure[]{
 \includegraphics[height=4cm,width=4cm]{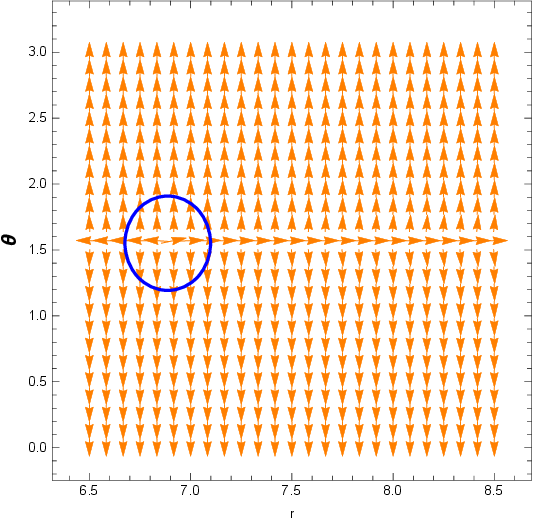}
 \label{10d}}
 \subfigure[]{
 \includegraphics[height=4cm,width=4cm]{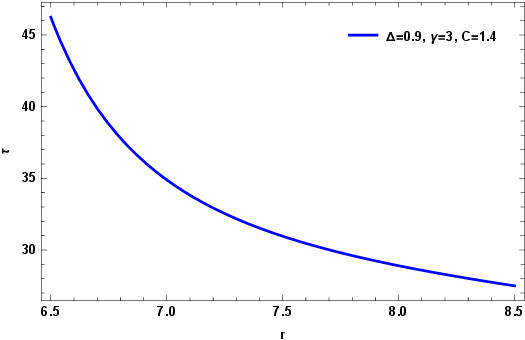}
 \label{10e}}
 \subfigure[]{
 \includegraphics[height=4cm,width=4cm]{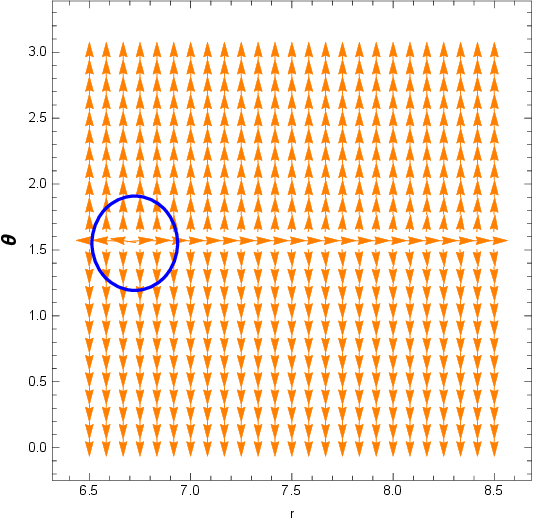}
 \label{10f}}
  \caption{\small{The curve described by Equation (\ref{RCT3}) is illustrated in Figures (\ref{10a}), (\ref{10c}), and (\ref{10e}). In Figures (\ref{10b}), (\ref{10d}), and (\ref{10f}), the zero points (ZPs) are located at coordinates $(r, \theta)$, corresponding to the nonextensive parameter $\Delta$.}}
 \label{m10}
 \end{center}
 \end{figure}

\section{Conclusion} \label{sec7}
In this paper, we studied the thermodynamic topology of AdS Einstein-power-Yang-Mills black holes, analyzing them through both the bulk-boundary and restricted phase space (RPS) frameworks. We investigate various non-extensive entropy models, including Barrow ($\delta$), Rényi ($\lambda$), Sharma-Mittal ($\beta$, $\alpha$), Kaniadakis ($K$), and Tsallis-Cirto ($\Delta$) entropy. Initially, we examine the thermodynamic topology within the bulk-boundary framework. Our findings reveal the significant impact of free parameters on topological charges. Notably, we observe two topological charges $(\omega = +1, -1)$ in relation to the non-extensive Barrow parameter and also when $\delta = 0$ in Bekenstein-Hawking entropy. For Rényi entropy, the topological charges vary with the value of $\lambda$, showing a transition from three topological charges $(\omega = +1, -1, +1)$ to a single topological charge $(\omega = +1)$ as $\lambda$ increases. Setting $\lambda$ to zero results in two topological charges $(\omega = +1, -1)$. Sharma-Mittal entropy displays three distinct ranges of topological charges influenced by $\alpha$ and $\beta$: when $\beta$ exceeds $\alpha$, we observe $(\omega = +1, -1, +1)$; when $\beta = \alpha$, we see $(\omega = +1, -1)$; and when $\alpha$ exceeds $\beta$, we encounter $(\omega = -1)$. Kaniadakis entropy shows variations in topological charges, with $(\omega = +1, -1)$ for any acceptable value of $K$, except when $K = 0$, where a single topological charge $(\omega = -1)$ is observed. For Tsallis-Cirto entropy, small $\Delta$ values result in $(\omega = +1)$, while increasing $\Delta$ to 0.9 leads to $(\omega = +1, -1)$.

A particularly intriguing aspect of this research is its application to the RPS framework. When we extend our analysis to this space using the specified entropies, we find that the topological charge consistently remains $(\omega = +1)$, regardless of the specific values of the free parameters for Rényi, Sharma-Mittal, and Tsallis-Cirto. This consistency suggests a stable topological structure within the RPS framework for these entropies. Additionally, for Barrow entropy in RPS, as $\delta$ increases from 0 to 0.8, the number of topological charges rises. This results in a more complex topological structure, indicating that the non-extensive parameter $\delta$ plays a crucial role in the behavior of topological charges. For Kaniadakis entropy, at small values of $K$, we observe $(\omega = +1)$. However, as the non-extensive parameter $K$ increases, we encounter different topological charges and classifications, resulting in $(\omega = +1, -1)$. This variation highlights the sensitivity of the topological structure to changes in the non-extensive parameter $K$.

These findings raise several intriguing questions for further exploration: How do these topological structures influence the physical properties of black holes? Can the stability observed in the RPS framework be leveraged to develop new theoretical models? What implications do these results have for our understanding of entropy in other complex systems? Exploring these questions could provide deeper insights into the nature of black holes and the role of non-extensive entropy in thermodynamic systems.\\
Moreover, the observed stability in the RPS framework across different entropy models suggests potential applications beyond black hole thermodynamics. Could this stability be a universal feature in other gravitational systems or even in non-gravitational contexts? How might these topological characteristics inform our understanding of phase transitions and critical phenomena in complex systems? The interplay between non-extensive parameters and topological charges opens up new avenues for research, potentially leading to a deeper comprehension of the fundamental principles governing entropy and topology in diverse physical systems. By addressing these questions, future research can build on our findings to explore the broader implications of thermodynamic topology, not only in the context of black holes but also in other areas of physics and beyond.

\begin{acknowledgments}
\.{I}.S. expresses gratitude to T\"{U}B\.{I}TAK, ANKOS, and SCOAP3 for their support. He is also thankful for the networking backing provided by COST Actions CA21106 and CA22113.
\end{acknowledgments}

\end{document}